\documentclass[manuscript]{aastex63}
\usepackage{bm}

\received{17 February 2020}
\revised{26 April 2020}
\accepted{14 May 2020}

\shorttitle{Formation of Haumea's Ring}
\shortauthors{Sumida et al.}

\begin{document}

\title{$N$-body Simulations of Ring Formation Process around the Dwarf Planet Haumea}

\correspondingauthor{Takanori Sasaki}
\email{takanori@kusastro.kyoto-u.ac.jp}

\author{Iori Sumida}
\affil{Department of Astronomy, Kyoto University, Kitashirakawa-Oiwake-cho, Sakyo-ku, Kyoto 606-8502, Japan}

\author[0000-0003-0745-2125]{Yuya Ishizawa}
\affil{Department of Astronomy, Kyoto University, Kitashirakawa-Oiwake-cho, Sakyo-ku, Kyoto 606-8502, Japan}

\author[0000-0002-6638-7223]{Natsuki Hosono}
\affiliation{Japan Agency for Marine-Earth Science and Technology, 2-15, Natsushima-cho, Yokosuka-city, Kanagawa 237-0061, Japan}
\affiliation{RIKEN Advanced Institute for Computational Science, 7-1-26 Minatojima-minami-machi, Chuo-ku, Kobe, Hyogo, Japan}

\author[0000-0003-1242-7290]{Takanori Sasaki}
\affil{Department of Astronomy, Kyoto University, Kitashirakawa-Oiwake-cho, Sakyo-ku, Kyoto 606-8502, Japan}


\begin{abstract}

Haumea is the only known trans-Neptunian object with a ring. The ring is in a position that causes a 3:1 spin-orbit resonance with the rotational period of Haumea, which has a triaxial shape. The non-axisymmetric gravitational field around Haumea is thought to affect the dynamics of the ring; however, the process of ring formation has not been elucidated. In this study, we analyze in some detail a potential ring formation scenario for Haumea. We first calculated the gravitational field around the triaxial ellipsoid and estimated the distance at which an object revolving around Haumea can exist stably using simulation that incorporated the time-varying gravitational field. The results of this simulation showed that the trajectory of the object became unstable just inside its current ring position. Next, we analytically derived the Roche radius for a rigid body revolving around a triaxial ellipsoid and showed that the Roche radius could be near the current ring position. Furthermore, as a parameter study, we performed $N$-body simulations using the coefficient of rubble pile restitution as a variable. Results demonstrated that, according to numerous parameters, the position of the Roche radius was near the current position of Haumea's ring.  Based on these findings, we can assume that there is a high possibility that the ring formed in the region between the boundary of the unstable region of the orbit and the Roche radius. The scenario presented in this study could help explain the process by which Haumea's ring formed.

\end{abstract}


\section{Introduction} \label{sec:Introduction}

Rings have been discovered around seven objects in the solar system. By the 1980s, rings had been discovered around all four giant planets. In the 21st century, rings had also been discovered around the Centaur asteroids Chariklo and Chiron, which orbit between Jupiter and Neptune \citep{2014Natur.508...72B, 2015A&A...576A..18O}. In 2017, a stellar occultation observed a ring around the dwarf planet Haumea \citep{2017Natur.550..219O}.

Haumea is one of five dwarf planets in the solar system and one of the trans-Neptunian objects (TNO) with an orbit outside Neptune. Haumea has two satellites \citep{2009AJ....137.4766R}. The rotation period of Haumea, which has a oblate triaxial ellipsoidal shape, is as fast as  3.9155 hours \citep{2010A&A...518L.147L}. The length of each semi-axis of Haumea is $a = 1161 \pm 30$ km, $b = 852 \pm 4$ km, and $c = 513 \pm 16$ km \citep{2017Natur.550..219O}. Haumea's ring is the first found around a celestial body with unambiguous triaxial ellipsoidal shape. It is narrow and dense with an optical thickness of 0.5 and a width of 70 km; it is coplanar with the equatorial plane of Haumea. The radius of the ring is $r = 2287^{+75}_{-45}$ km, which is where a 3:1 spin-orbit resonance ratio occurs between the revolution of the ring and the rotational period of Haumea ($a_{3:1} = 2285 \pm 8 $ km) \citep{2017Natur.550..219O}; in other words, while the particles that make up the ring rotate one period, Haumea itself rotates three periods.

Haumea's ring is thought to be affected by the non-axisymmetric gravitational field caused by Haumea's triaxial shape, but the physical mechanism that determines the position of the ring is poorly understood. The purpose of this study is to explain why Haumea's ring is at a 3:1 mean kinetic resonance position.

\citet{2012MNRAS.419.2315O} showed that Haumea experienced fission due to its rapid rotation and that stripped fragments could have turned into its satellites or the Haumea family. Considering the fact that the current angular momentum of the system, including Haumea and its satellites, is extremely high, a previous study used $N$-body simulations to show how Haumea was accelerated and split due to the collision of small objects. In general, orbits of objects outside the corotation radius of a planet evolve due to tidal forces and move further outward. The corotation radius of Haumea is
\begin{equation}
	r_{\mathrm{cr}} = \left(\frac{GM_{\mathrm{Haumea}}}{\Omega^2_{\mathrm{Haumea}}} \right)^{1/3} = 1104 \; \mathrm{km},
\end{equation}
where $M_{\mathrm{Haumea}} = 4.006 \times 10^{21}$ kg is the mass of Haumea and $\Omega_{\mathrm{Haumea}} = 4.4575 \times 10^{−4}$ rad s$^{-1}$ is the angular velocity of its rotation. As Haumea's longest semi-axis, $a = 1161$ km, is outside its corotation radius, any fragments ejected by splitting are also located outside this radius.

\begin{figure*}
  \plotone{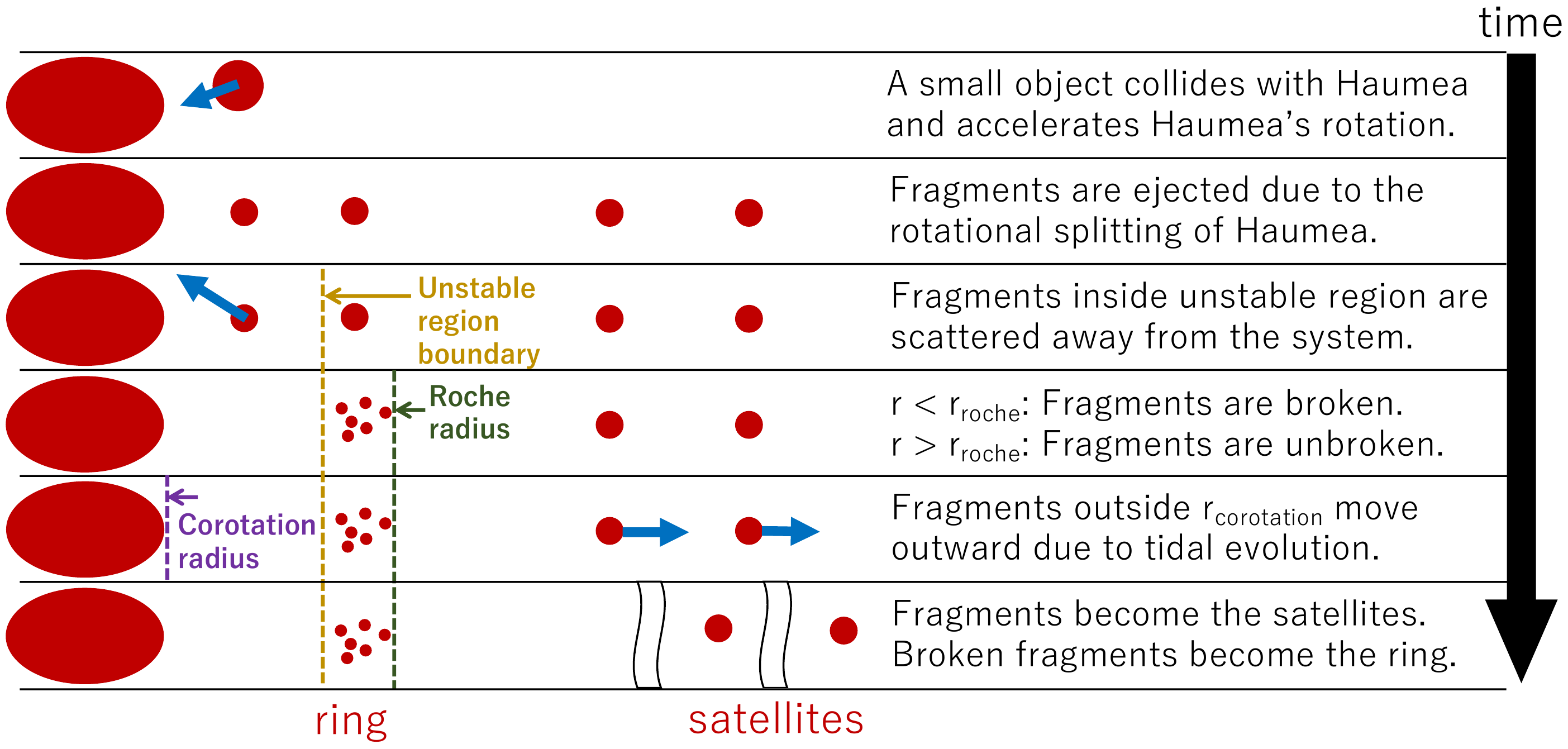}
  \caption{Schematic of one scenario for the formation of the Haumea system. Wave lines in the last step of the process are scale break lines those indicate significant radius gaps.}
  \label{fig:Scenario}
\end{figure*}

In this study, modelled after \citet{2012MNRAS.419.2315O}, a ring formation scenario for Haumea is proposed (Fig.\ref{fig:Scenario}). The mechanism of ring formation through rotational fission was one of the possible scenarios mentioned by \citet{2014Natur.508...72B} for the origin of Chariklo's ring, but it was not investigated. Here we analyze the scenario in detail for Haumea's ring. We consider a situation in which fragments were ejected due to the splitting of Haumea. The ejected fragments behave differently depending on their locations. Near Haumea, the non-axisymmetric component of the gravitational field of triaxial ellipsoidal Haumea is large, which produces an unstable region where the object cannot stably revolve around Haumea. Among the objects in the stable region, those inside the Roche radius are destroyed by tidal force, and the remaining objects then turn into a ring by orbiting Haumea. On the other hand, those ejected fragments found outside the Roche radius move further outward to become the satellites.

\citet{2019NatAs...3..146S} showed that among particles around a triaxial ellipsoidal object, those existing around a Lindblad resonance all undergo orbital evolution under torque. They numerically calculated the trajectories of particles subjected to a triaxial ellipsoidal gravity field. When the central object was Haumea, it was shown that on a 10-year time scale, all particles inside the 2:1 spin-orbit resonance disappeared. Furthermore, it was observed that particles slightly outside the 2:1 spin-orbit resonance were pushed out further with time. However, the boundaries of the final unstable region were not shown. The position of the 3:1 orbital resonance, where the ring of Haumea exists, is located outside the 2:1 orbital resonance. We aim to find out how close the boundaries of the unstable region of the particles that orbit Haumea are to the 3:1 orbital resonance. 

The Roche radius has been calculated analytically and numerically  when the central celestial body is a sphere, and the object revolving around the central celestial body is a rigid body or a perfect fluid, the expression of the Roche radius can be analytically written down \citep[e.g.,][]{1963ApJ...138.1182C}. \citet{2012MNRAS.424.1419L} performed an $N$-body simulation to find the location of the Roche radius of a rubble-pile object revolving around a spherical central object. \citet{2001Icar..149..375D} derived the expressions for the Roche radius for solid biaxial ellipsoids. However, the calculation of the Roche radius has not yet been performed for a triaxial ellipsoidal central object.

In this study, the position of the unstable region boundary and the position of the Roche radius around the triaxial ellipsoidal Haumea are calculated by $N$-body calculation to verify the formation process of Haumea's ring.

\section{Orbital Unstable Region around Haumea} \label{sec:calculation_1}

We investigate the boundary of the orbital unstable region around the triaxial shaped ellipsoid that is Haumea. An object is affected by a non-axisymmetric gravitational field when in the vicinity of an elliptically shaped celestial body. In this section, the position of the unstable region boundary of the orbit is obtained by examining the stability of the motion of particles revolving around Haumea. Although \citet{2019NatAs...3..146S} have already executed a similar experiment, there are some differences in calculation methods between our study and theirs. We use an analytic expression for the gravitational potential instead of the potential expansion used in theirs. Moreover, while they included Stokes-like friction, which simulates collisions between particles, our study does not have it and only includes the gravitational field of the central object.

\subsection{Methods of Calculation}

The motion of a particle revolving around Haumea is calculated by the following equation;
\begin{equation}
	\frac{\mathrm{d}^2\bm{r_i}}{\mathrm{d}t^2} = -\nabla U(\bm{r_i}),
\end{equation}
where $U$ is the gravitational potential of the triaxial ellipsoid (Haumea) rotating at $T_\mathrm{Haumea} = 3.9155$ hours. It is assumed that Haumea has a uniform density $\rho = 1.885$ g cm$^{-3}$, and is a triaxial ellipsoid with axes radii of $a = 1161 \pm 30$ km, $b = 852 \pm 4$ km, $c = 513 \pm 16$ km. Haumea's axis of rotation is the shortest axis ($c$-axis). See Appendix \ref{sec:appendix_a} for the detailed formula of the gravitational potential $U$ outside the triaxial ellipsoid. In this calculation, the force acting on the particles only considers the gravitational field of Haumea. However, if the particle size is as small as $\mu$m, the radiation pressure of sunlight cannot be ignored and may affect the location of the unstable region \citep[e.g.,][]{2018MNRAS.479.4560K}.

The numerical integration of the equation of motion uses the second-order Leap Frog method. In this calculation, particles revolving around Haumea behave as massless test particles. That is, there is no interaction between the particles, and Haumea is not affected by the accreting particles or the reaction of gravity. Particles that fall on Haumea are removed from the calculation without affecting the body of Haumea.

The number of particles that orbit Haumea is $N = 50000$. The particles are arranged on the same plane as the equatorial plane of Haumea (plane parallel to the $a$-axis and $b$-axis), and the area density distribution is uniform. The inner edge of the initial major radius of the orbit is $a_{\mathrm{in}} = a$ (the longest axis radius of Haumea) and the outer edge is $a_{\mathrm{max}} = 3$ $r_{\mathrm{ring}}$. The initial values of the orbital eccentricity and the orbital inclination are all 0. Each particle is given a Kepler velocity, $v_{\mathrm{Kepler}}$, in a forward direction with respect to the rotation of Haumea, as shown below:
\begin{equation}
	v_{\mathrm{Kepler}} = \sqrt{\frac{GM_{\mathrm{Haumea}}}{a_{\mathrm{ptcl}}}},
\end{equation}
where $a_{\mathrm{ptcl}}$ is the initial major orbital radius of the particle.

\subsection{Numerical Results}

\begin{figure*}
  \gridline{\fig{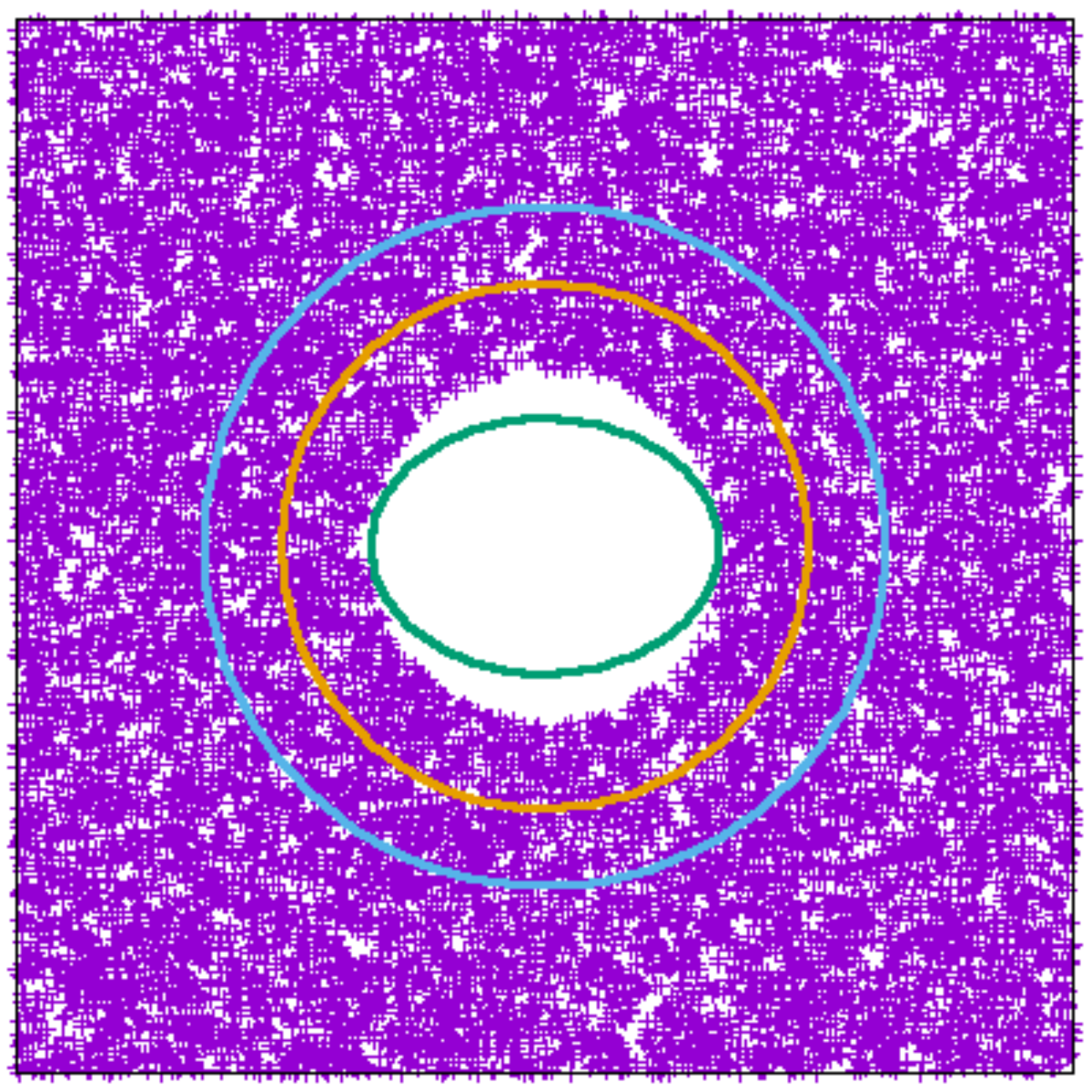}{0.3\textwidth}{$t = 0$ day}
            \fig{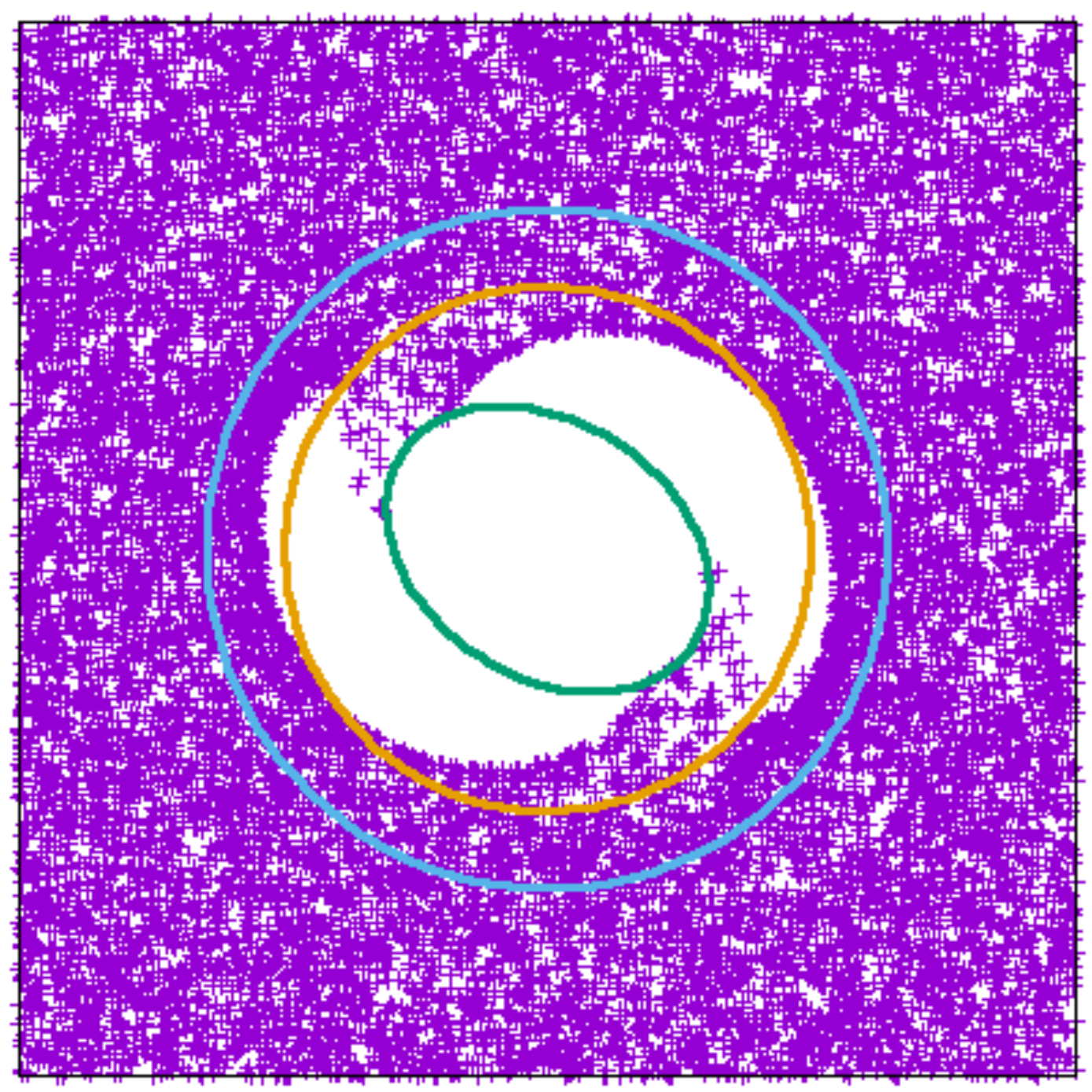}{0.3\textwidth}{$t = 1$ day}
            }
  \gridline{\fig{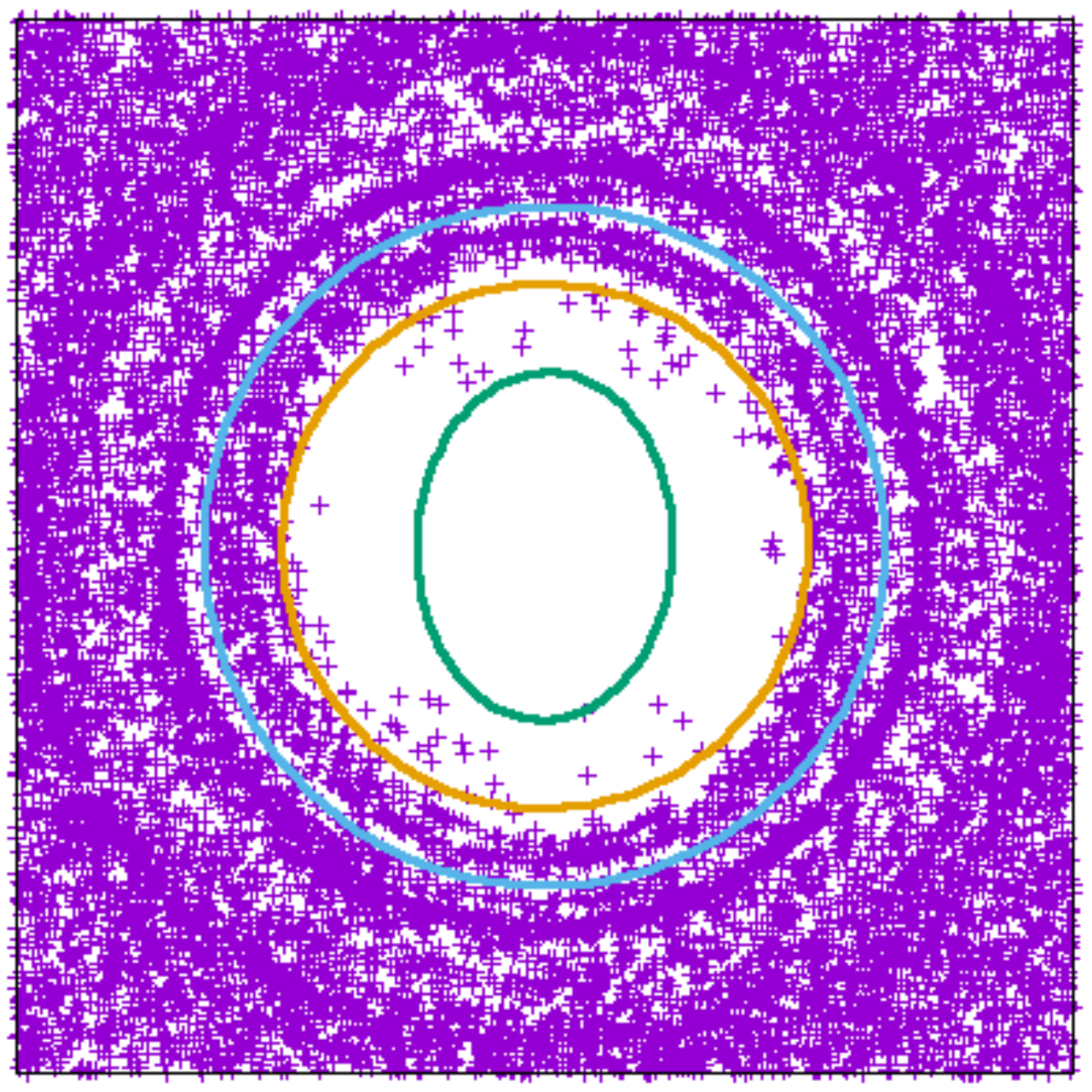}{0.3\textwidth}{$t = 10$ days}
            \fig{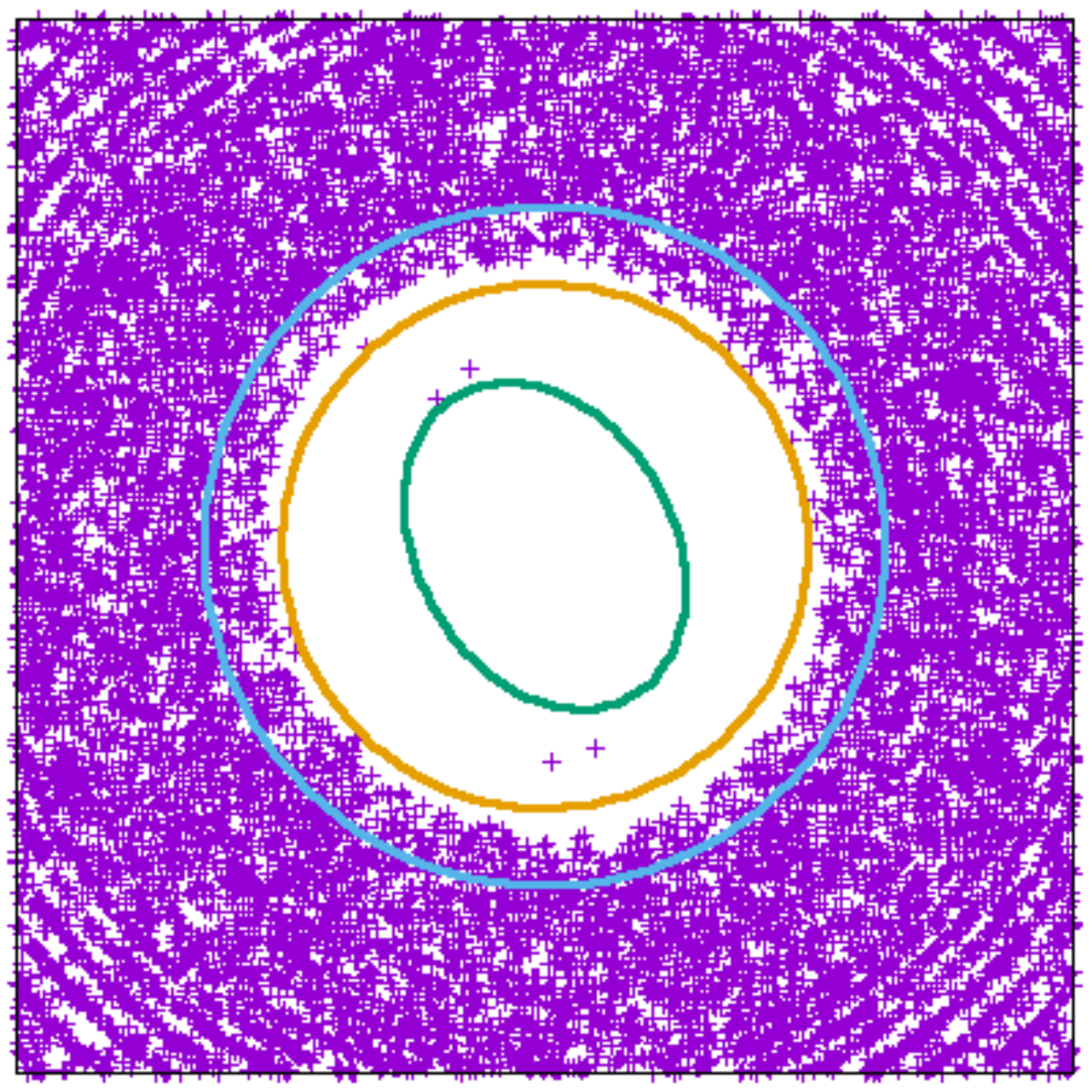}{0.3\textwidth}{$t = 100$ days}
            }
  \gridline{\fig{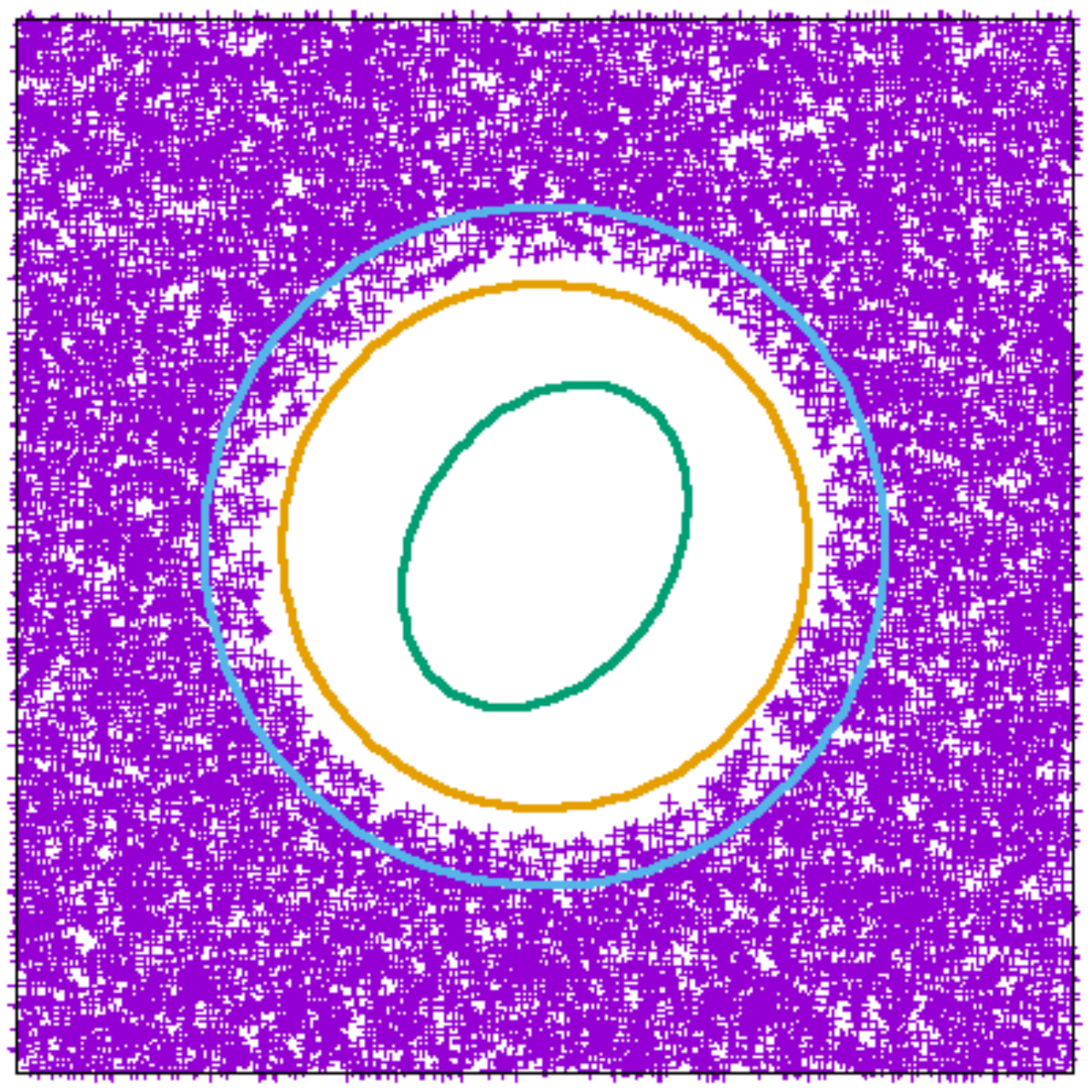}{0.3\textwidth}{$t = 1000$ days}
            \fig{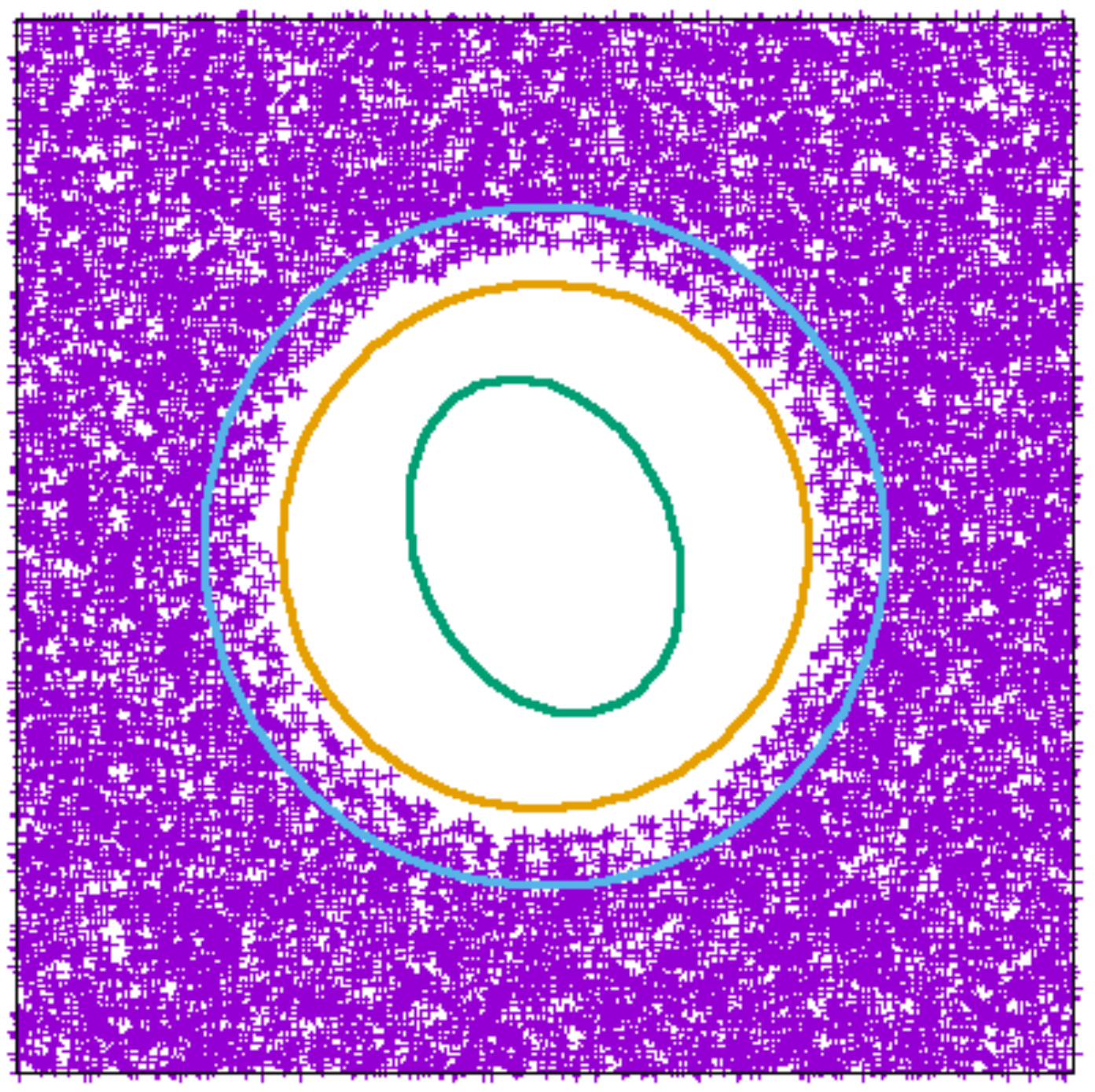}{0.3\textwidth}{$t = 10000$ days}
            }
  \caption{Time variation of the distribution of particles around Haumea. The distributions after 0, 1, 10, 100, 1000, 10000 days from the initial condition are shown. The center green ellipse is Haumea, the outer yellow circle is the 2:1 spin-orbit resonance position ($a_{2:1} = 0.767$ $r_{\mathrm{ring}}$), and the light blue circle is the 3:1 spin-orbit resonance position ($a_{3:1} =  1.0$ $r_{\mathrm{ring}}$).
  \label{fig:unstablexy}}
\end{figure*}

\begin{figure*}
  \gridline{\fig{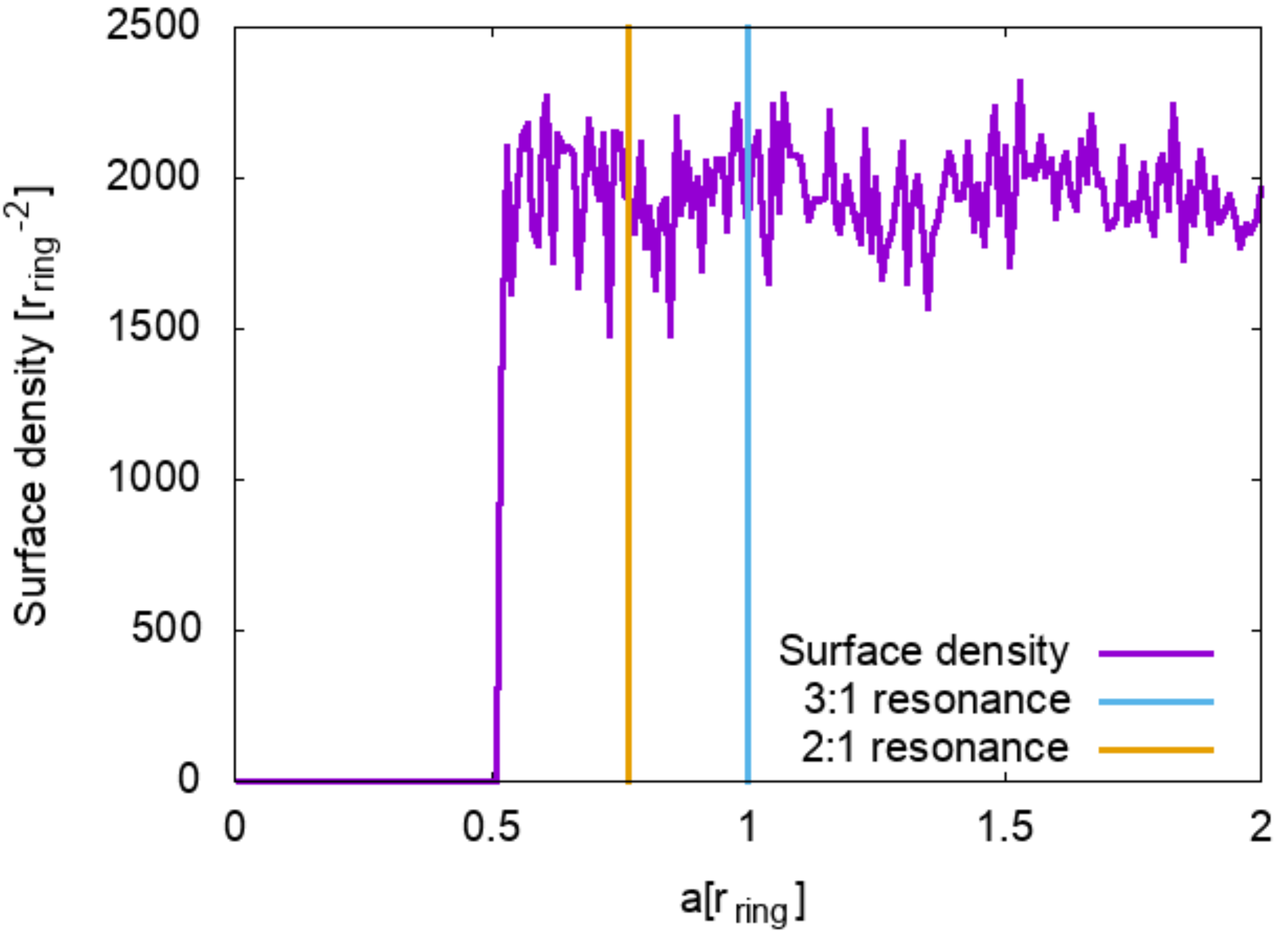}{0.4\textwidth}{$t = 0$ day}
            \fig{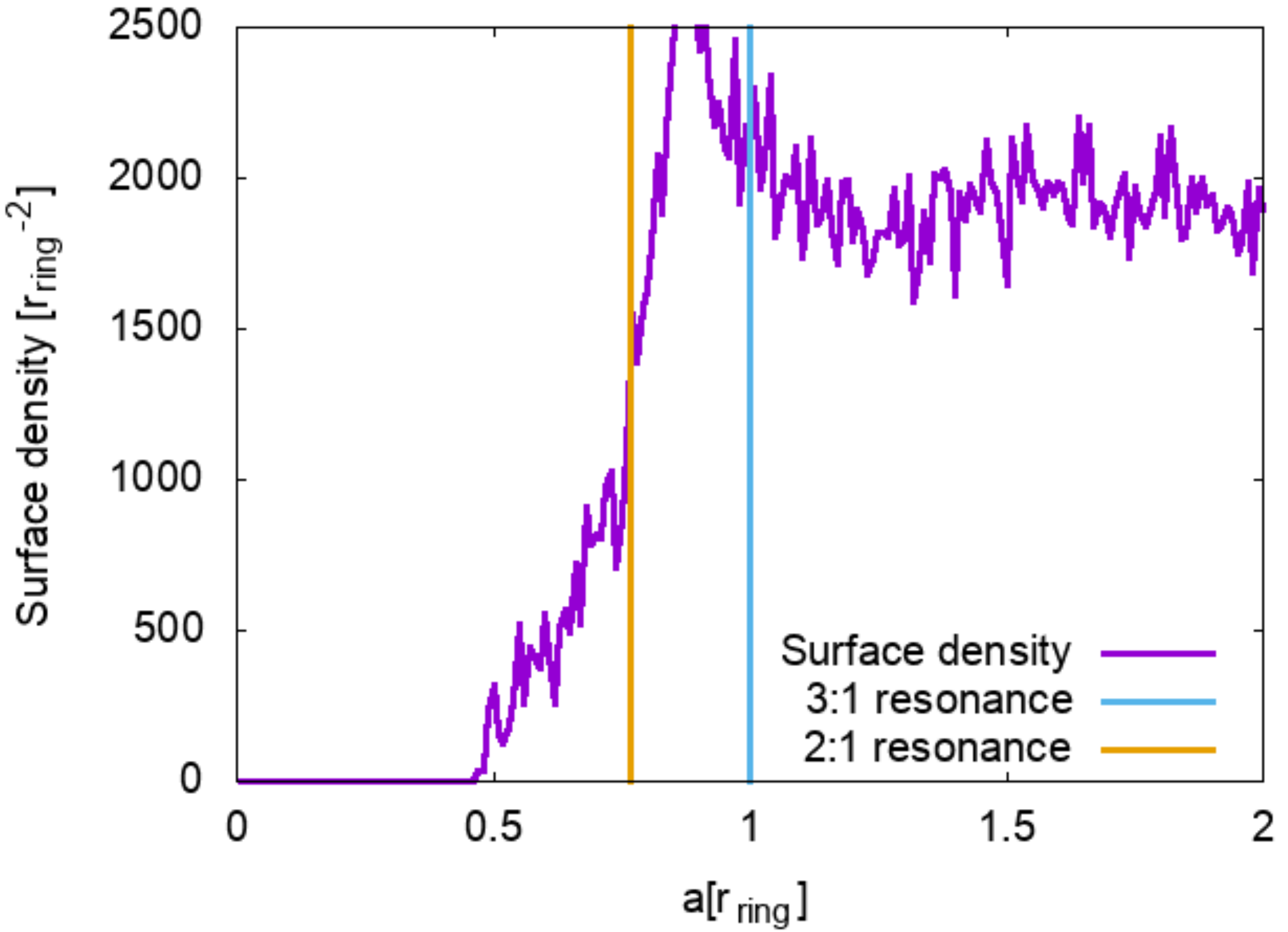}{0.4\textwidth}{$t = 1$ day}
            }
  \gridline{\fig{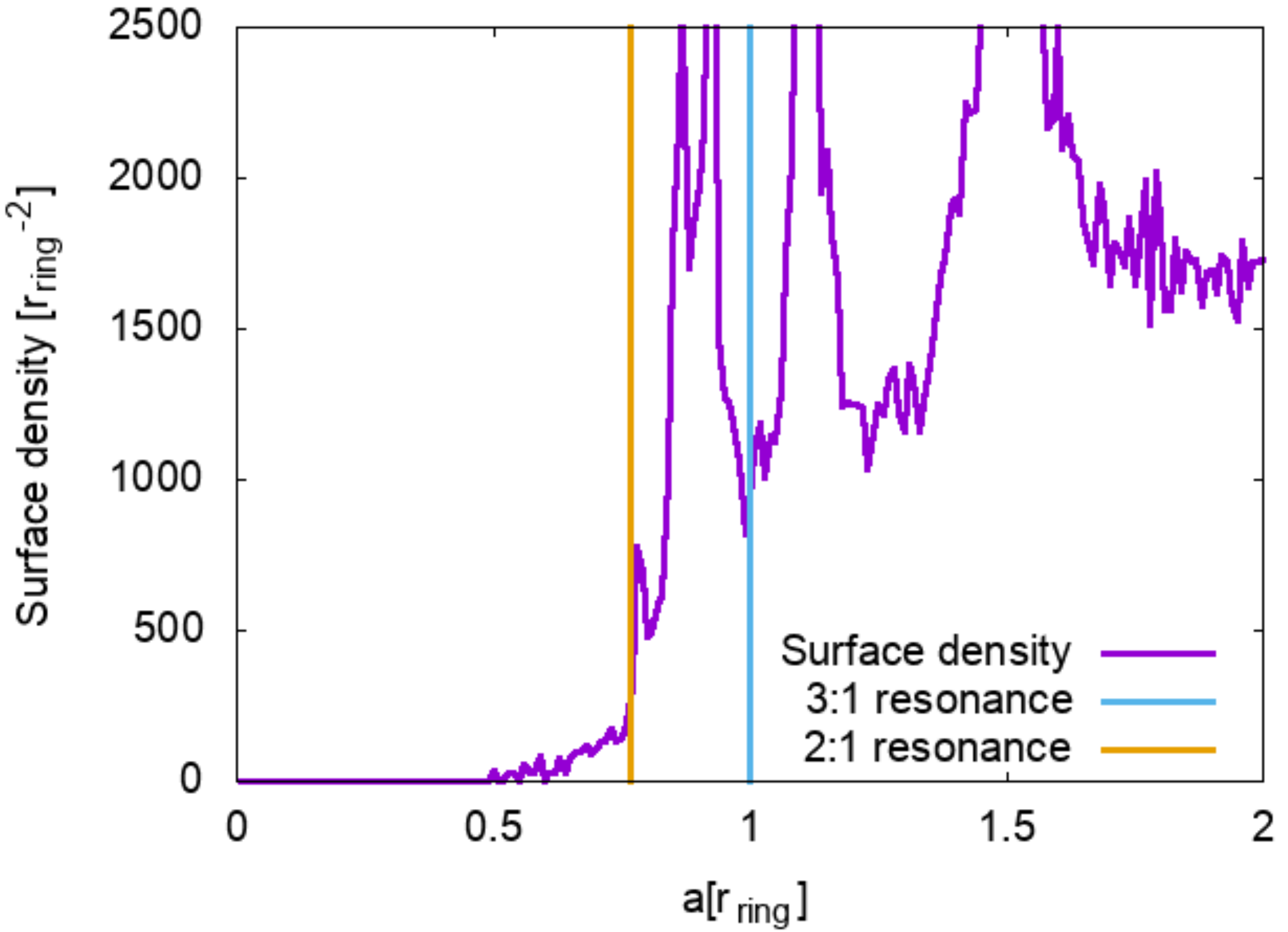}{0.4\textwidth}{$t = 10$ days}
            \fig{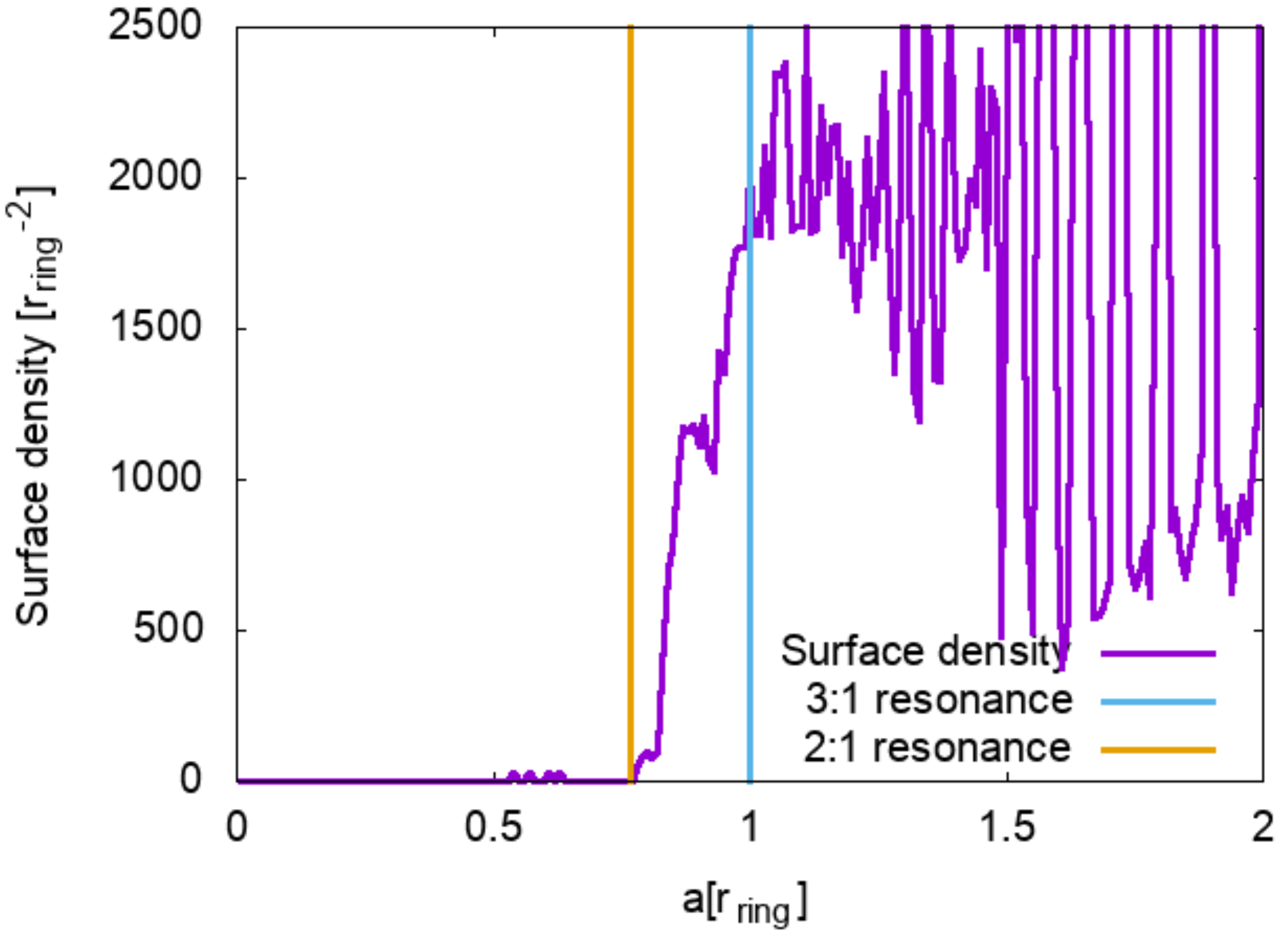}{0.4\textwidth}{$t = 100$ days}
            }
  \gridline{\fig{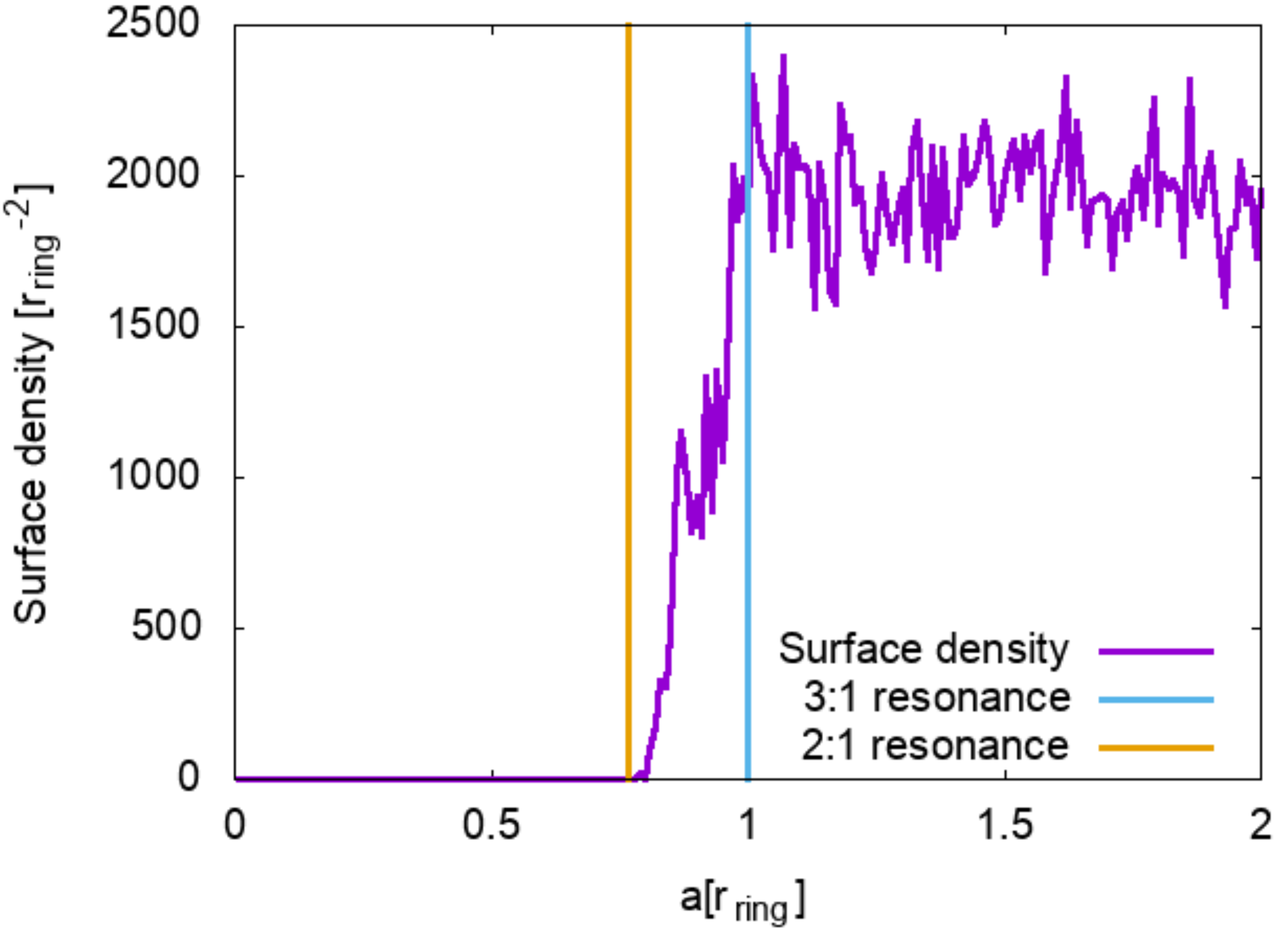}{0.4\textwidth}{$t = 1000$ days}
            \fig{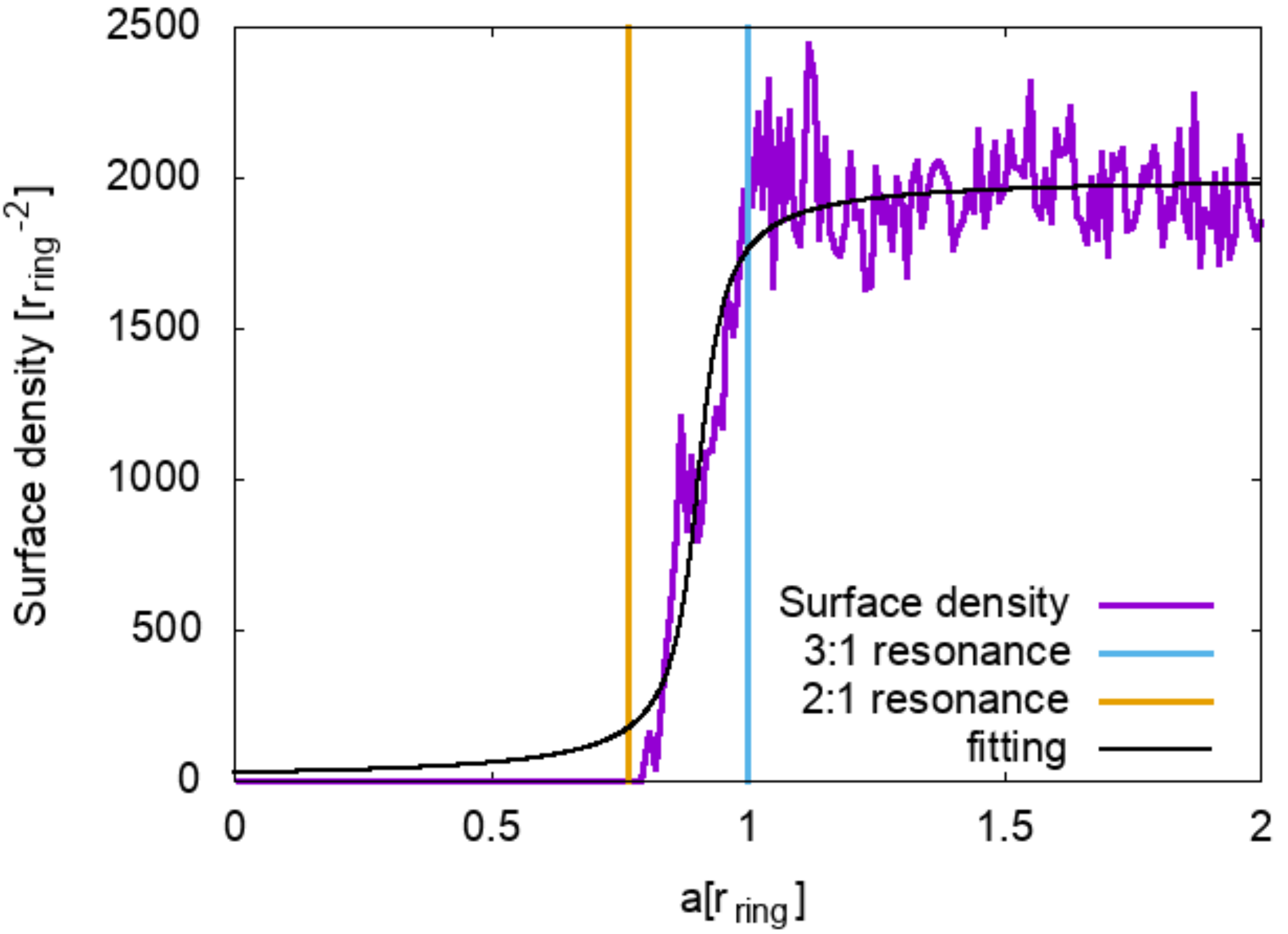}{0.4\textwidth}{$t = 10000$ days}
            }
  \caption{Time change of distribution of areal density of particles (number per unit area) around Haumea. The distribution after 0, 1, 10, 100, 1000, 10000 days from the initial condition are shown. The position of the 2:1 orbital resonance and the position of the 3:1 orbital resonance are also shown. The figure at $t =$ 10000 days shows an approximate curve of the area density distribution.
  \label{fig:unstablesigma}}
\end{figure*}

Figure \ref{fig:unstablexy} shows the particle distribution at time $t$ = 0, 1, 10, 100, 1000, and 10000 days. Figure \ref{fig:unstablesigma} shows the radial distribution of the areal density (number of particles per unit area) at the same times. At $t$ = 1 day, a particle-free region was formed near Haumea. This was caused by particles near Haumea moving outward or falling into Haumea under strong positive or negative torque from a non-axisymmetric gravitational field. At $t$ = 10 days and $t$ = 100 days, there were regions in which the particle density distribution oscillated in the radial direction. This was caused by the propagation of the density wave generated near the 2:1 orbital resonance ($a = a_{2:1}$) by the torque of the non-axisymmetric gravitational field to the outside. After $t$ = 100 days, almost no particles existed inside $a_{2:1}$. After $t$ = 1000 days, the mean value of the area density distribution outside $a_{3:1}$ stabilized and there was hardly any temporal evolution in the distribution. 

In Fig. \ref{fig:unstablesigma}, the final state---the surface density distribution at $t$ = 10000 days---was fitted with a smooth function. The orbit unstable region boundary $a_{\mathrm{UR}}$ is defined as the position where the surface density is half the average surface density in the outer region. According to our result, $a_{\mathrm{UR}} = 0.905$ $r_{\mathrm{ring}}$. This orbital unstable region boundary exists between the position of the 2:1 orbital resonance and the position of the 3:1 orbital resonance ($a_{2:1}<a_{\mathrm{UR}}<a_{3:1}$), which is consistent with the results in \citet{2019NatAs...3..146S}.

\subsection{Discussions of the Numerical Result}

A similar, but more thorough analysis of the general stability of orbits around a nonspherical body in rotation, starting with Keplerian orbits was done with quite a lot of detail in \citet{2004P&SS...52..685H} in which they used a more general (although approximate) expression for the gravity potential than what is presented in this paper. For Haumea, there are indications that it must be differentiated, with a denser core than the crust. Since the gravity potential used in this study is valid for a homogeneous ellipsoid, it is possible that the position of the unstable region boundary can change when differentiation is taken into account. These aspects were taken care of in \citet{2004P&SS...52..685H}. If we consider Haumea has a uniform density and use the rotational angular velocity and mass of the Haumea, we can calculate a mass-distribution parameter $\sigma$ defined in the paper,
\begin{equation}
	\sigma = \frac{I_{yy}-I_{xx}}{I_{zz}-I_{xx}} = 0.57,
\end{equation}
where $I_{xx}$, $I_{yy}$ and $I_{zz}$ are the principal moments of inertia. \citet{2004P&SS...52..685H} showed unstable region for the cases $\sigma = 1$ and $0.5$. For $\sigma = 1$, the unstable region boundary is equivalent to $0.87$ $r_{\mathrm{ring}}$. For $\sigma = 0.5$, there are two stable-unstable region boundaries, and the outer one is equivalent to $0.90$ $r_{\mathrm{ring}}$. These values are almost consistent with our result ($a_{\mathrm{UR}} = 0.905$ $r_{\mathrm{ring}}$).

In the rotational fission scenario, the speeds and positions of the particles and fragments ejected from the main body do not correspond to those of Keplerian orbits. The integration of the equations of motion should be used as initial conditions those of the output from fission models. In fact, \citet{2012MNRAS.419.2315O} showed that at least a percentage of the fragments have speeds larger than the escape velocity for Haumea, in the collision-induced rotational breakup scenario. In future studies, it would be necessary to simulate the outcome of a disruption and then integrate the orbital evolution of the fragments to discuss the ring formation process more realistically.

We also note that the surface density obtained by our study does not have sharp boundaries, while the Haumea's ring has sharp boundaries and is much narrower than what we obtained. The difference may be due to the oversimplifying assumptions of our model; self-gravity, dynamical friction and collisions would play important roles in the dynamics of large numbers of particles. For instance, \citet{2016ApJ...821...18P} showed that self-gravity can counteract the orbital precession caused by the $J2$ component of the gravity potential and that has been used to place constraints on the mass of the ring for Chariklo.

\section{Roche Radius around Haumea} \label{sec:calculation_2}

In this section, we investigate the Roche radius surrounding Haumea. The Roche radius is the distance within which objects bounded only by their own gravity (hereinafter ``companions'') can reach the central celestial body without tidal destruction. We calculate the location of the Roche radius around Haumea using analytical and $N$-body calculations.

\subsection{Analytical Estimation of Roche Radius}

We can analytically derive the Roche radius when the central celestial body is a triaxial ellipsoid. In instances where the central object is a sphere, an analytical derivation of the Roche radius has been previously established \citep[e.g.,][]{1963ApJ...138.1182C}. When the central celestial body is a triaxial ellipsoid, it is necessary to derive the magnitude of the tidal force acting on the companion from the gravitational field expression of the triaxial ellipsoid.

First, we derive the Roche radius when the companion orbiting the central celestial body is a rigid body. A rigid body is an object that is not deformed by external forces such as tidal and centrifugal forces. In this case, Haumea's ring is on its equatorial plane (a plane parallel to the $a$- and $b$-axes), and the tidal force on the equatorial plane is greatest on the longest $a$-axis. The Roche radius around a triaxial inequality ellipsoid is not spherical, and is probably shaped like a triaxial ellipsoid. The longest axis of that triaxial ellipsoid is the same as the longest axis of the central body. Objects orbiting around Haumea in a circular orbit move in and out of the Roche radius. Once within the Roche radius, we assume that the object will soon be tidally disrupted and will not reassemble. Therefore, the Roche radius on the longest axis of the Haumea can be considered an ``effective'' Roche limit. It is also assumed that the rotation of the companion is synchronized with the orbit of the tidal force.

For purposes of this study, let the mass of the central object be $M$, the density be $\rho_M$, the companion mass be $m$, the radius be $r$, and the density be $\rho_m$. The companion is located at $(x, 0, 0)$ on $a$-axis. The companion's self-gravity $\mathcal{F}_G$ is:
\begin{equation}
	\mathcal{F}_G = \frac{Gm}{r^2} = \frac{4}{3}\pi G \rho_m r.
\end{equation}
Since the rotation of the companion is synchronized with the revolution, the rotation angular velocity $\omega$ is:
\begin{equation}
	\omega = \sqrt{\frac{GM}{x^3}},
\end{equation}
so that the centrifugal force $\mathcal{F}_C$ due to the rotation of the companion is:
\begin{equation}
	\mathcal{F}_C = r\omega^2 = \frac{GMr}{x^3} = \frac{4 \pi G a b c \rho_M r}{3x^3}.
\end{equation}
The tidal force $\mathcal{F}_T$ acting on the companion is described in the following equation (see Appendix \ref{sec:appendix_b} for detail derivation):
\begin{equation}
	\mathcal{F}_T = -\frac{4 \pi G abc\rho_M r}{(a^2 - c^2)^{3/2} k^2} \left( F(x)-E(x) + x \frac{d}{dx}(F(x)-E(x))\right).
\end{equation}

\begin{figure}
  \plotone{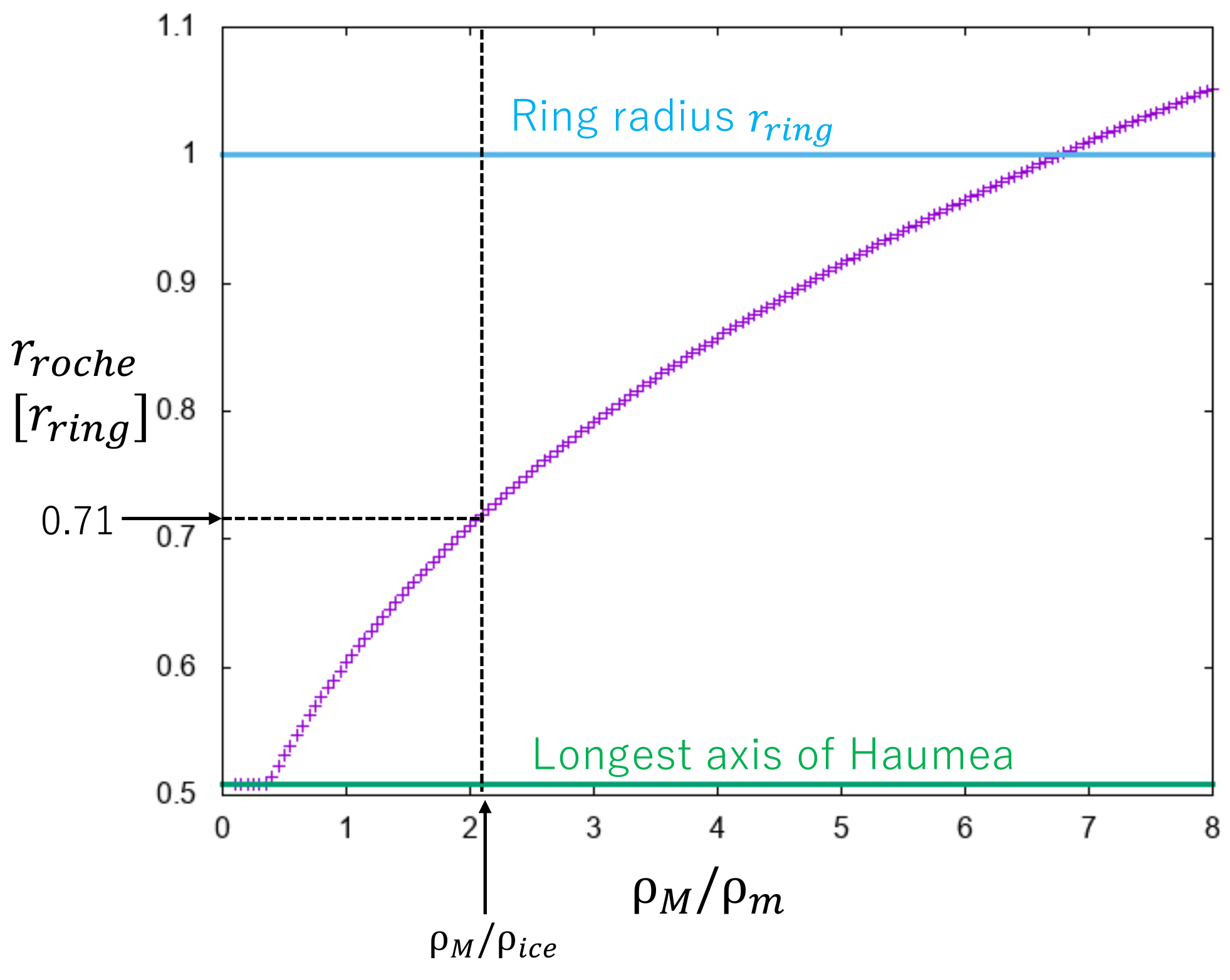}
  \caption{Density ratio dependence of the rigid Roche radius around a triaxial ellipsoid. $r = 0.5081$ $r_{\mathrm{ring}}$ is Haumea's longest axis radius. When the density of Haumea is $\rho_M = 1.885$ g cm$^{-3}$ and the density of the object is $\rho_m = \rho_{\mathrm{ice}} = 0.913$ g cm$^{-3}$, the Roche radius is rroche, $r_{\mathrm{roche,rigid}}=0.71$ $r_{\mathrm{ring}}$.}
  \label{fig:rocherigidanaly}
\end{figure}

The tidal force, from the central celestial body acting on the companion, and the rotational centrifugal force are the forces that try to destroy the companion. On the other hand, the self-gravity acting on the companion is the force that suppresses companion destruction. Since these forces balance at the Roche radius, we can derive the rigid body Roche radius using:
\begin{equation}
	\mathcal{F}_G = \mathcal{F}_C + \mathcal{F}_T.
\end{equation}

The solution to this equation depends on the density ratio $\rho_M/\rho_m$ between the central object and the companion. Therefore, the Roche radius was determined as a function of the density ratio (Fig. \ref{fig:rocherigidanaly}). While observations revealed the density of Haumea, $\rho_M = 1.885$ g cm$^{-3}$, the density of the objects that make up the ring are unknown. For this reason, the density of the companion was assumed to be equal to the density of the ice $\rho_{\mathrm{ice}} = 0.913$ g cm$^{-3}$. The resulting Roche radius is $r_{\mathrm{roche,rigid}}=0.71$ $r_{\mathrm{ring}}$.

We note that the derived expression of the Roche limit for a rigid body around an ellipsoidal body is not a general one, but only valid when the companion has its rotation locked to the orbital period and for systems in which the companion's mass would be negligible compared with the central object's mass.

Next, we considered an instance in which the companion orbiting the central celestial body is a perfect fluid. The perfect fluid is one that has no viscosity, and with a shape deformed such that the boundary surface runs parallel to the equipotential surface. However, because of the tidal force that the central celestial body exerts on the companion---the self-gravity of which depends on its shape---it is difficult to derive the Roche radius analytically as is the case in a rigid body scenario. Therefore, we use the expression for the case in which Haumea is approximated to be a sphere of the same volume. Assuming that the rotation of the companion is synchronized with the revolution, the Roche radius for the spherical central object is derived analytically \citep[e.g.,][]{1963ApJ...138.1182C}:
\begin{equation}
	r_{\mathrm{roche,fluid}}=2.455R\left(\frac{\rho_M}{\rho_m}\right)^{1/3}=1.09\; r_{\mathrm{ring}},
\end{equation}
where $R = (abc)^{1/3}$.

From the above, the relationship between the rigid body roche radius, the perfect fluid body roche radius, and the ring radius is:
\begin{equation}
	r_{\mathrm{roche,rigid}} < r_{\mathrm{ring}} < r_{\mathrm{roche,fluid}}.
\end{equation}

In addition, it is possible to restrict the position of the actual roche radius from the positions of both a rigid body roche radius and a perfect fluid roche radius. As a real object behaves like a viscous fluid, it exhibits an intermediate property between a rigid body and a perfect fluid body; in other words, the roche radius for a real object exists somewhere between $r_{\mathrm{roche,rigid}} = 0.71$ $r_{\mathrm{ring}}$ and $r_{\mathrm{roche,fluid}} = 1.09$ $r_{\mathrm{ring}}$. This is consistent with our scenario in which the Haumea's ring formed just inside the Roche radius.

As the roche radius for an existing viscous fluid cannot be derived analytically, the actual roche radius can be estimated using the numerical calculation described in the following section.

\subsection{Method of $N$-body Simulation}

We calculate the distance at which the companion orbiting Haumea will be destroyed by the tide when approaching Haumea. In this numerical calculation, the companion is assumed to be a rubble pile object. A rubble pile object is composed of many particles interacting with each other, and the motion of each particle is calculated by $N$-body calculation.

The motion of each particle is calculated by the following equation: 
\begin{equation}
	\frac{d^2\bm{r_i}}{dt^2}=-\nabla U(\bm{r_i}) - \sum_{i \neq j} Gm_j \frac{\bm{r_j}-\bm{r_i}}{|\bm{r_j}-\bm{r_i}|^3},
\end{equation}
where $U$ is the gravitational potential of Haumea. The numerical integration of the equation of motion uses the second-order Leap Frog method. In the usual method for calculating the interaction between any two particles, $N$-body calculation requires a large computational cost of $\mathcal{O}(N^2)$ when the number of particles is $N$.  Here, we used a library called the “Framework for Developing Particle Simulator” (FDPS) to simulate the large-scale parallel particle method \citep{2016PASJ...68...54I}. This improved the computational cost to $\mathcal{O}(N \log N)$.

The number of particles that make up the rubble pile are $N$ = 1000. The mass of the entire rubble pile is $10^{-5}M_{\mathrm{Haumea}}$. Each particle is a hard sphere and does not penetrate or break. For simplicity, we assumed that each particle had the same size and mass. When particles collide with each other, a repulsive force and a frictional force are both present. The relative velocities of the two particles before and after the collision are:
\begin{equation}
	\bm{v_n'}=-\epsilon_n \bm{v_n}
\end{equation}
\begin{equation}
	\bm{v_t'}=\epsilon_t \bm{v_t}
\end{equation}
where $\bm{v}$ is the relative velocity before the collision, $\bm{v'}$ is the relative velocity after the collision, and the subscripts $n$ and $t$ represent the normal and tangential component, respectively. The restitution coefficient parameters $\epsilon_n$ and $\epsilon_t$ can have values from 0 to 1. It has been experimentally found that the actual values of $\epsilon_n$ and $\epsilon_t$ vary depending on conditions such as the size of the object, the collision speed, and the collision angle \citep[e.g.,][]{1998Icar..133..310H}. In our simulation, the value of the coefficient of restitution was changed when $\epsilon_n$, $\epsilon_t$ = 0.0, 0.2, 0.4, 0.6, 0.8, 1.0, and a subsequent parameter study was performed. For simplicity, we did not consider the dependence of $\epsilon_n$ and $\epsilon_t$ on the collision velocity. Further, both the rotation of the particles and the fixing force between them were ignored. A collision between two particles was detected when the distance between them was smaller than the sum of their radii. At collision, we reset the distance between two particles such that this distance is equal to the sum of the radii of the two particles (see \citet{2019ApJ...885..132I} in detail).

\subsection{Numerical Calculation of Roche Radius}

The initial shape of the rubble pile object is spherical and each constituent particle is arranged such that they have a close-packed structure. The packing ratio of particles in the close-packed structure is 74.0\%. In the initial state, the average density of the rubble pile is set to that of ice  $\rho_{\mathrm{ice}} = 0.913$ g cm$^{-3}$. In other words, the density of the constituent particles is $\rho_{\mathrm{ptcl}} = \rho_{\mathrm{ice}} / 0.740 = 1.23$ g cm$^{-3}$. The initial position of the rubble pile object was located at various distances from Haumea on its equatorial plane. The initial eccentricity of the rubble pile object is 0, and it revolves at the Kepler speed in a direction that goes forward with respect to the rotation of Haumea. The initial spin rate of the rubble pile object set equal to the orbital rotation of the object. The trajectory calculation continues until the rubble pile orbits 20 times around Haumea or until tidal disruption. When the mass of the rubble pile object becomes less than half of its initial mass, it is considered destroyed by tidal force.

\begin{figure*}
  \gridline{\fig{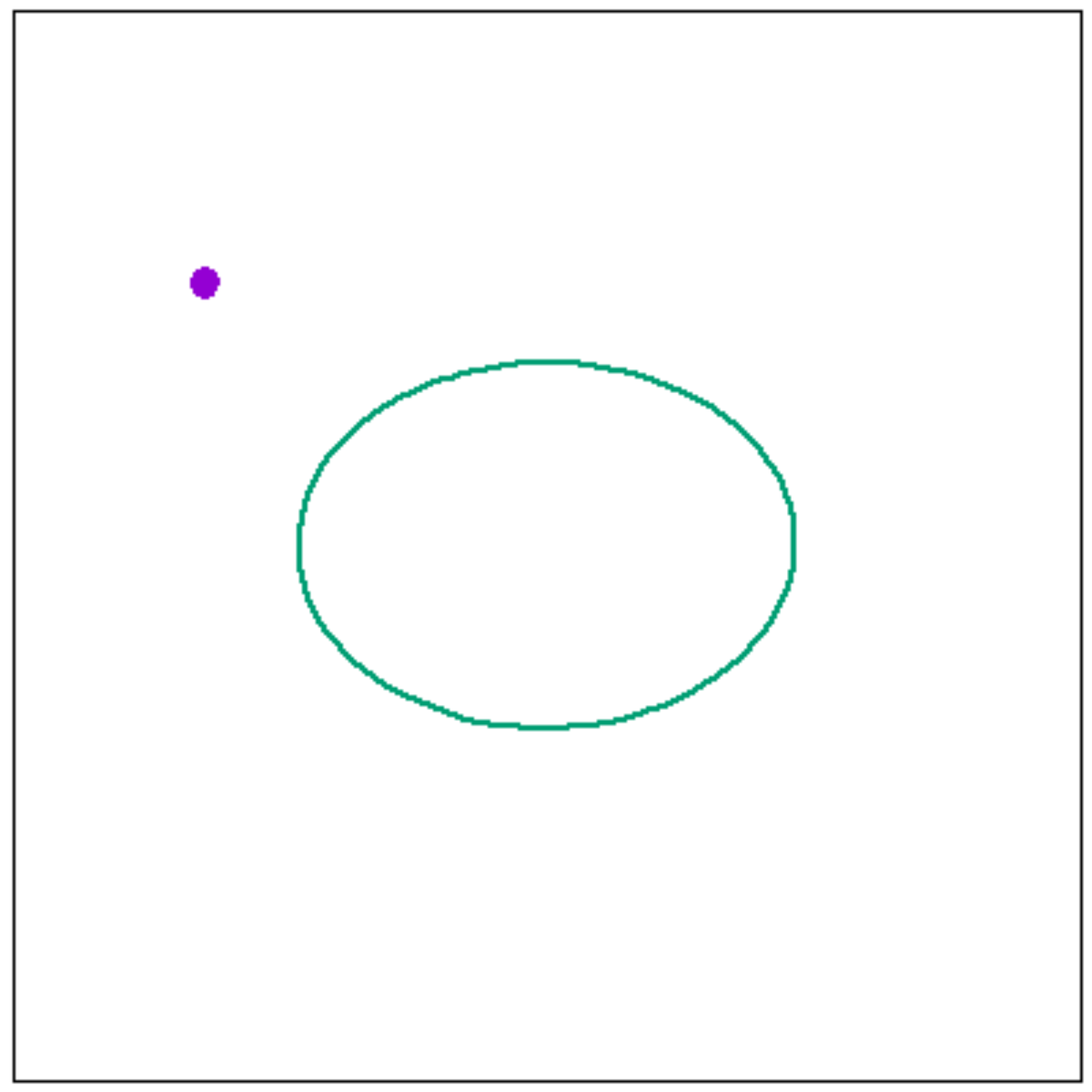}{0.3\textwidth}{$t = 0.0$ hour}
            \fig{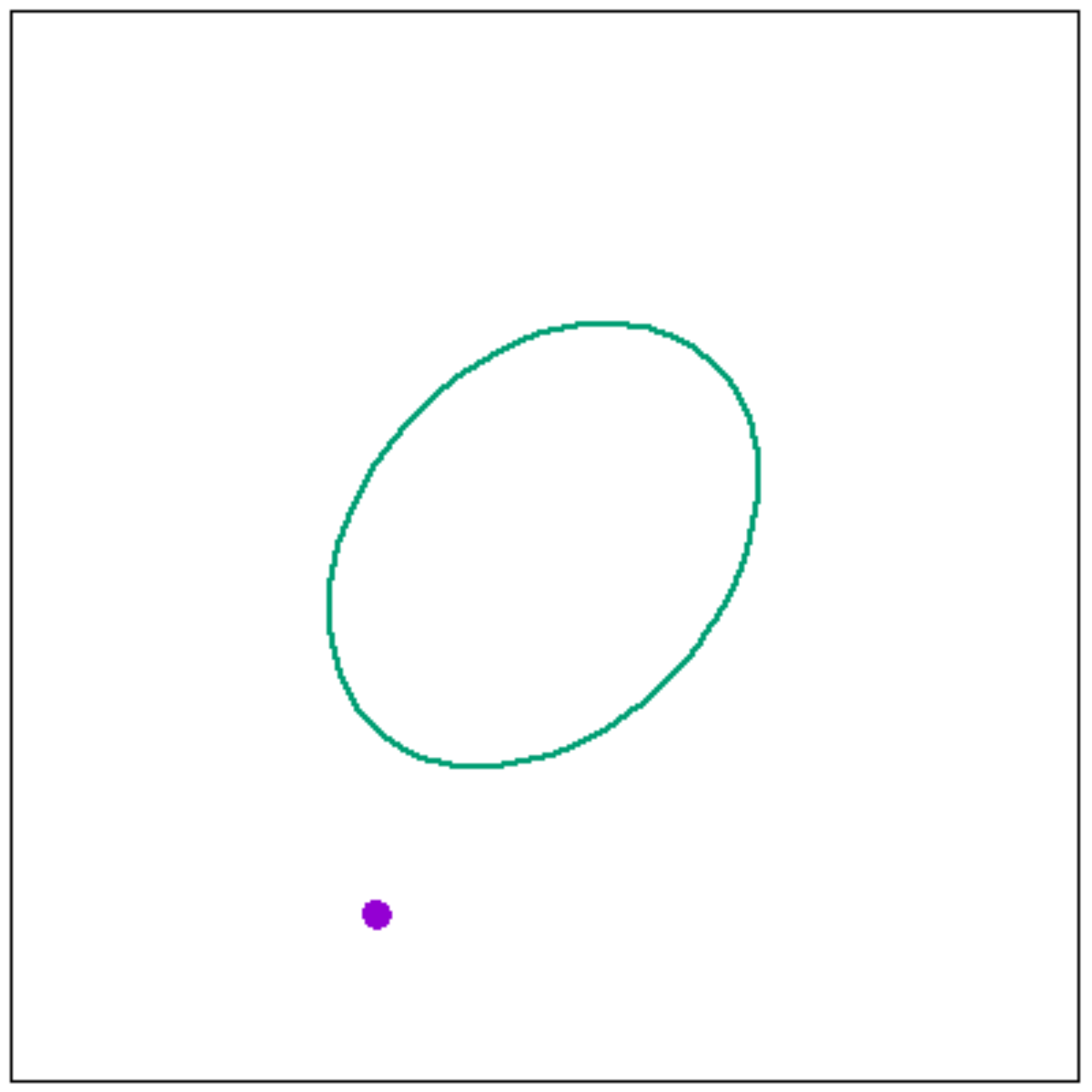}{0.3\textwidth}{$t = 4.4$ hours}
            \fig{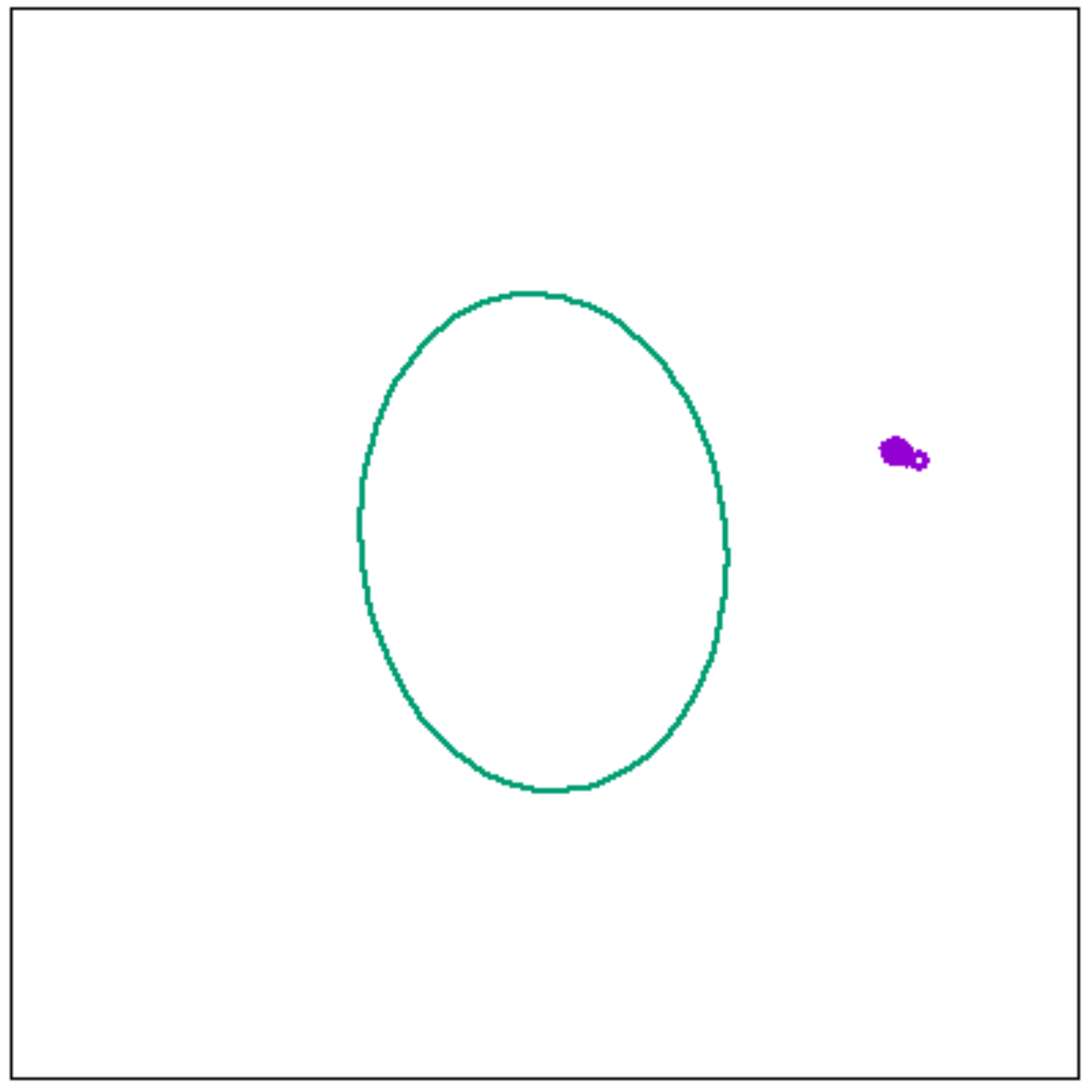}{0.3\textwidth}{$t = 8.9$ hours}
            }
  \gridline{\fig{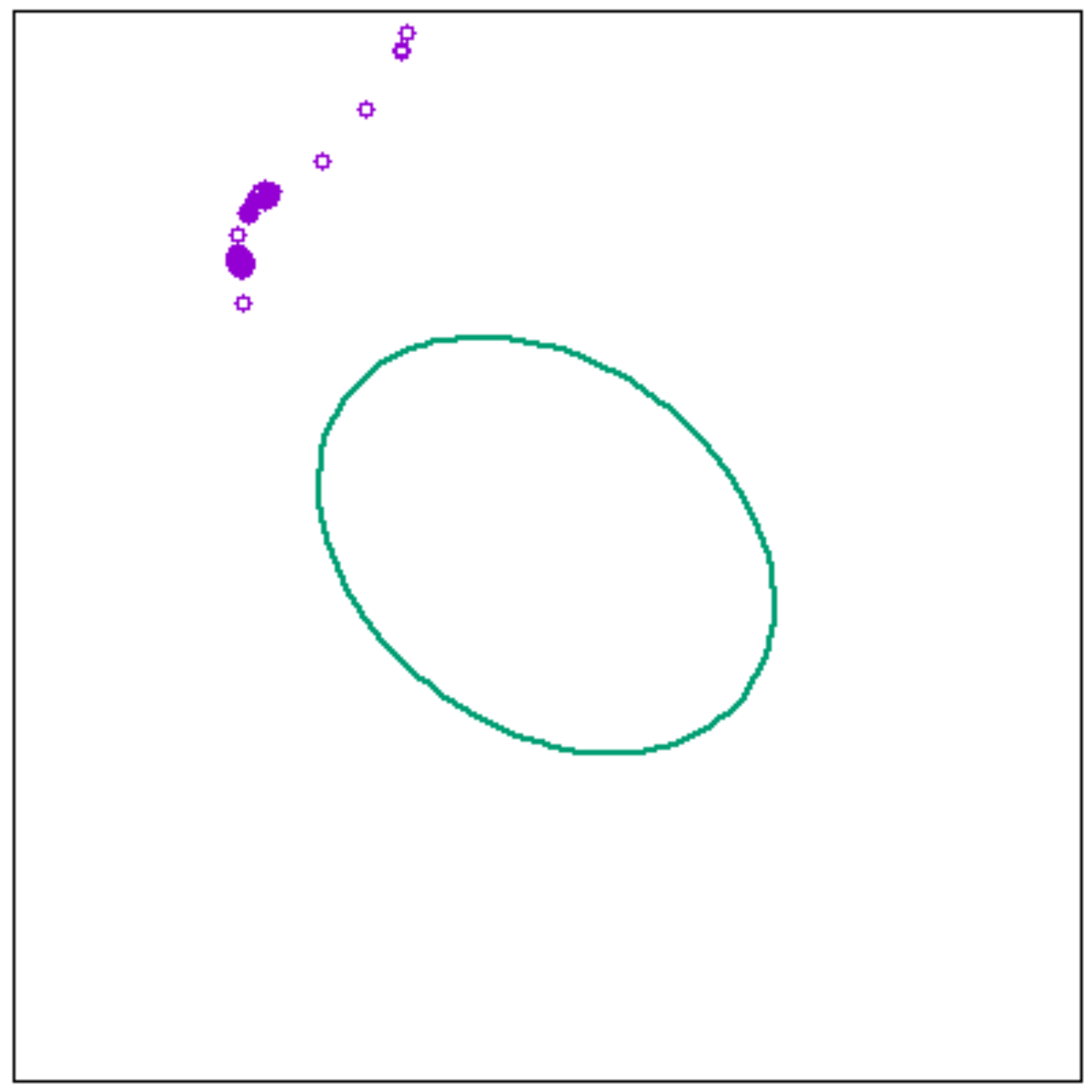}{0.3\textwidth}{$t = 13.3$ hours}
            \fig{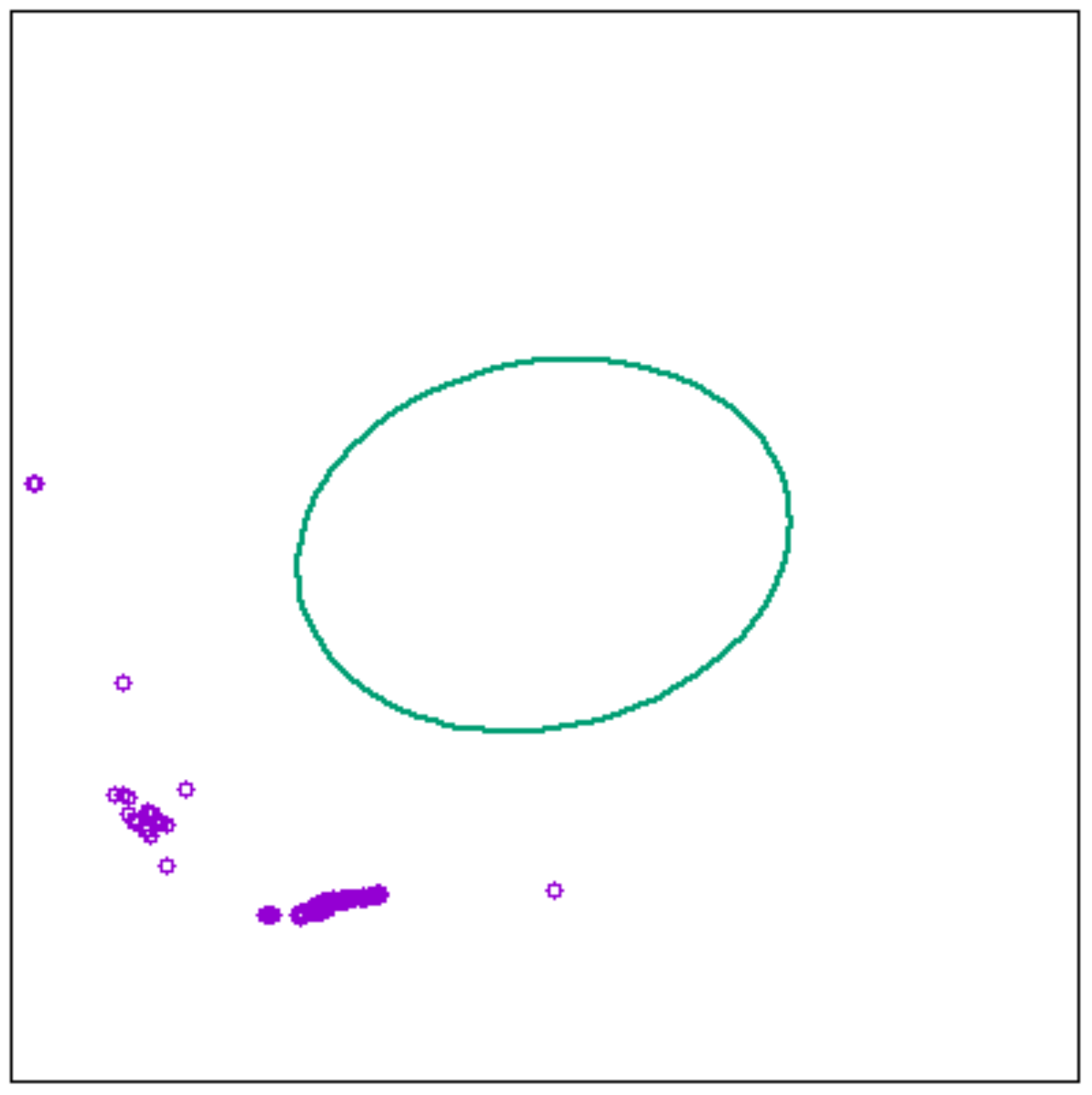}{0.3\textwidth}{$t = 17.7$ hours}
            \fig{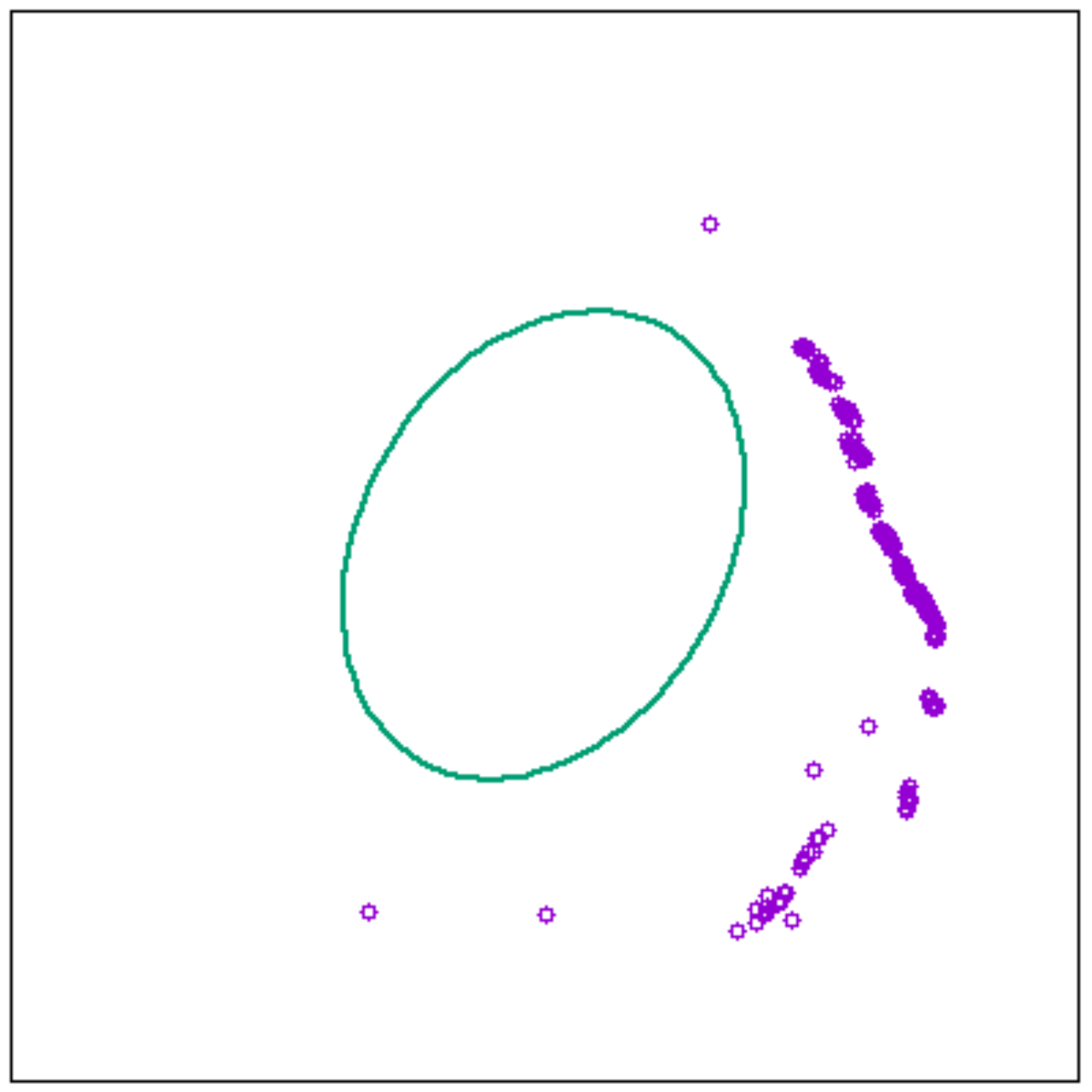}{0.3\textwidth}{$t = 22.2$ hours}
            }
  \caption{Example of tidal break (Run 2 in Table \ref{table:ifdisrupt}). The projection of the behavior of Haumea and a rubble pile object revolving around it onto the $x-y$ plane.
  \label{fig:rochedisruptorbit}}
\end{figure*}

\begin{figure*}
  \gridline{\fig{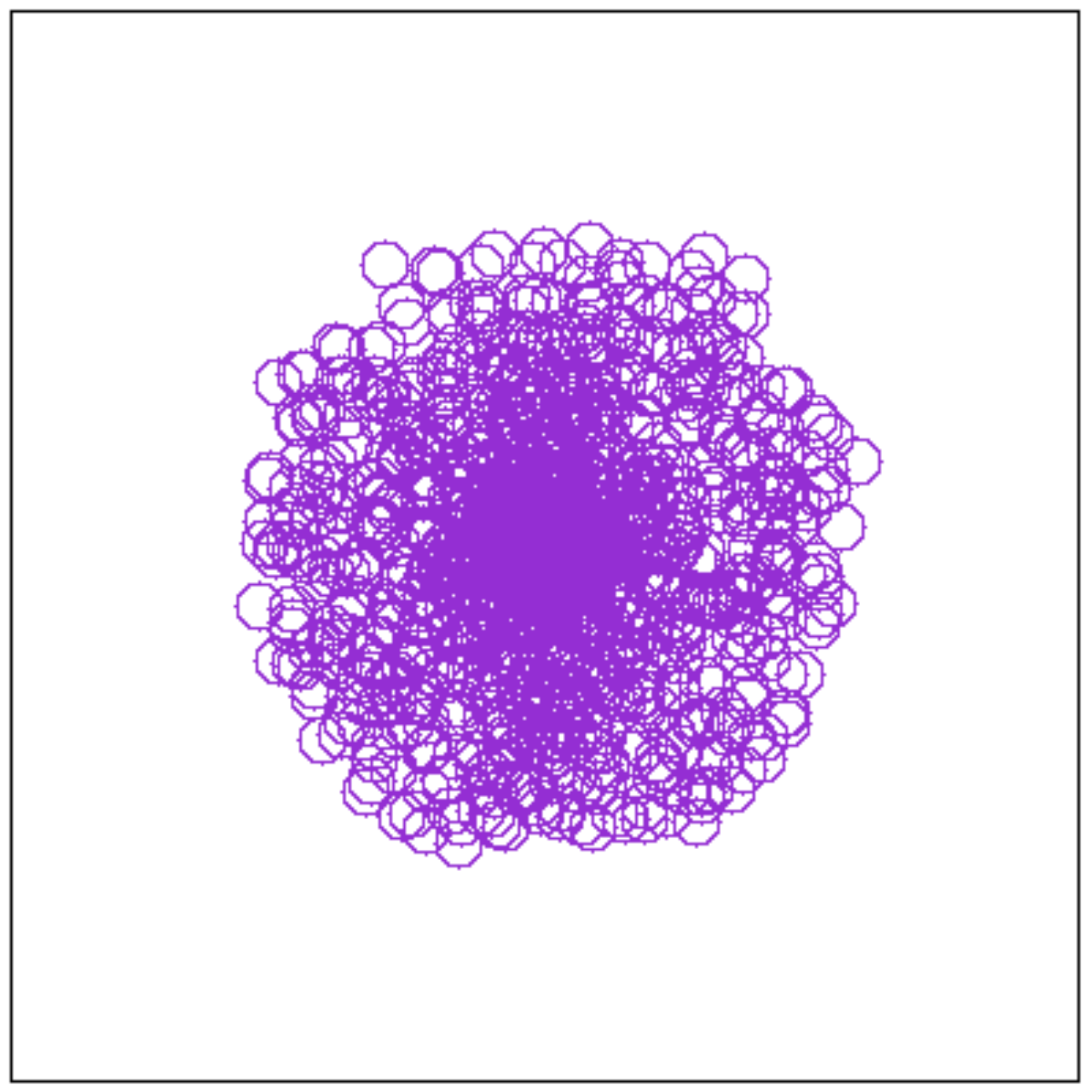}{0.3\textwidth}{$t = 0.0$ hour}
            \fig{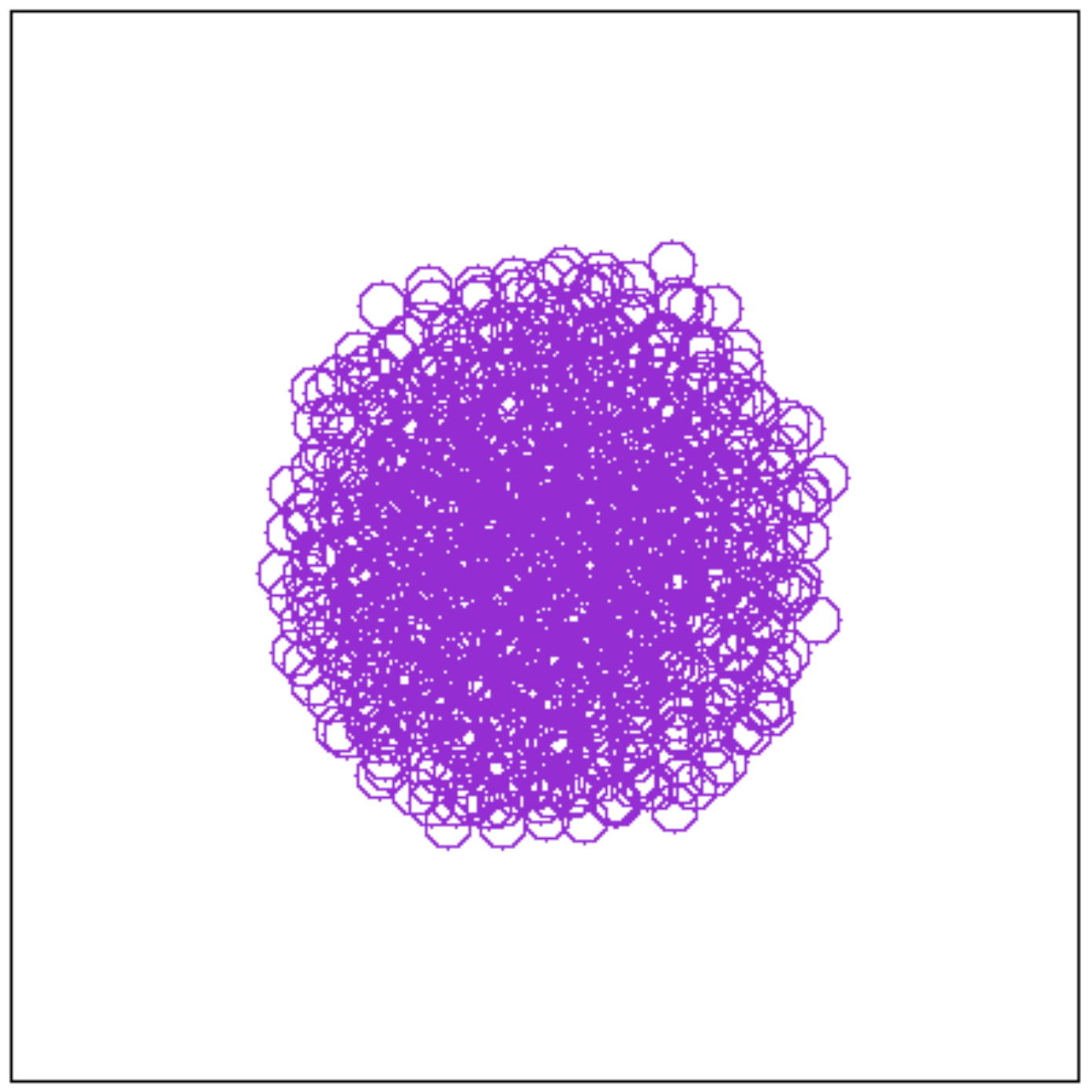}{0.3\textwidth}{$t = 4.4$ hours}
            \fig{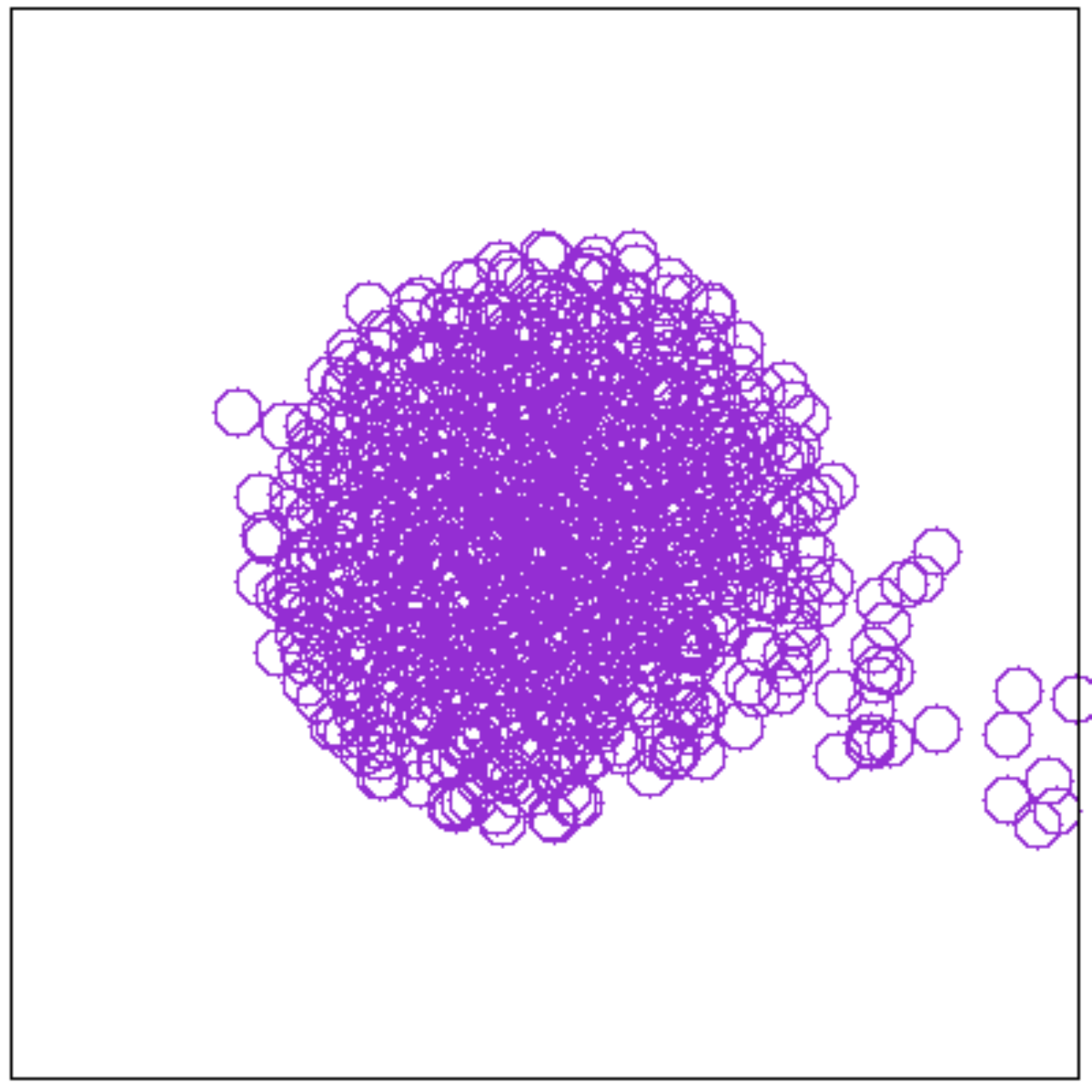}{0.3\textwidth}{$t = 8.9$ hours}
            }
  \gridline{\fig{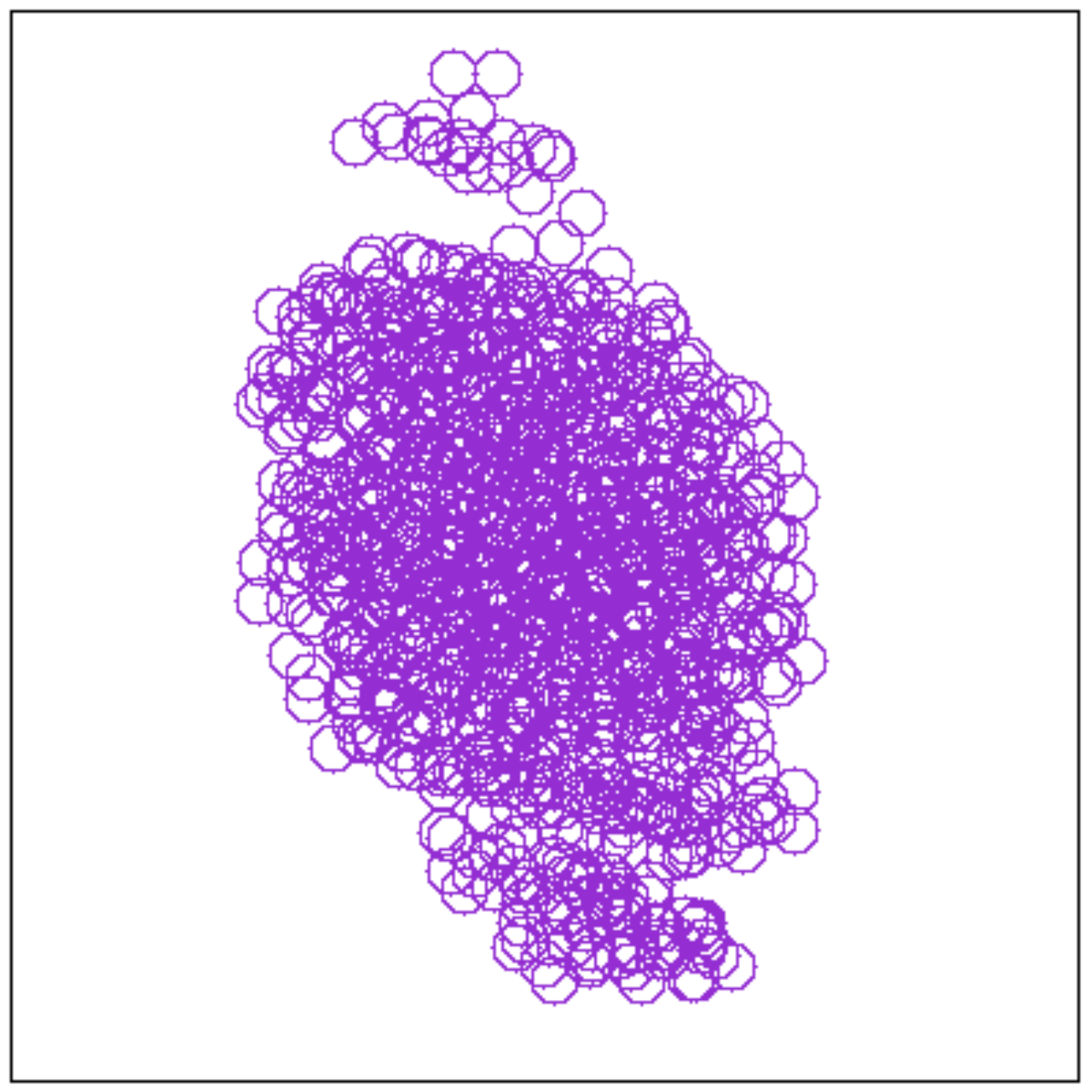}{0.3\textwidth}{$t = 13.3$ hours}
            \fig{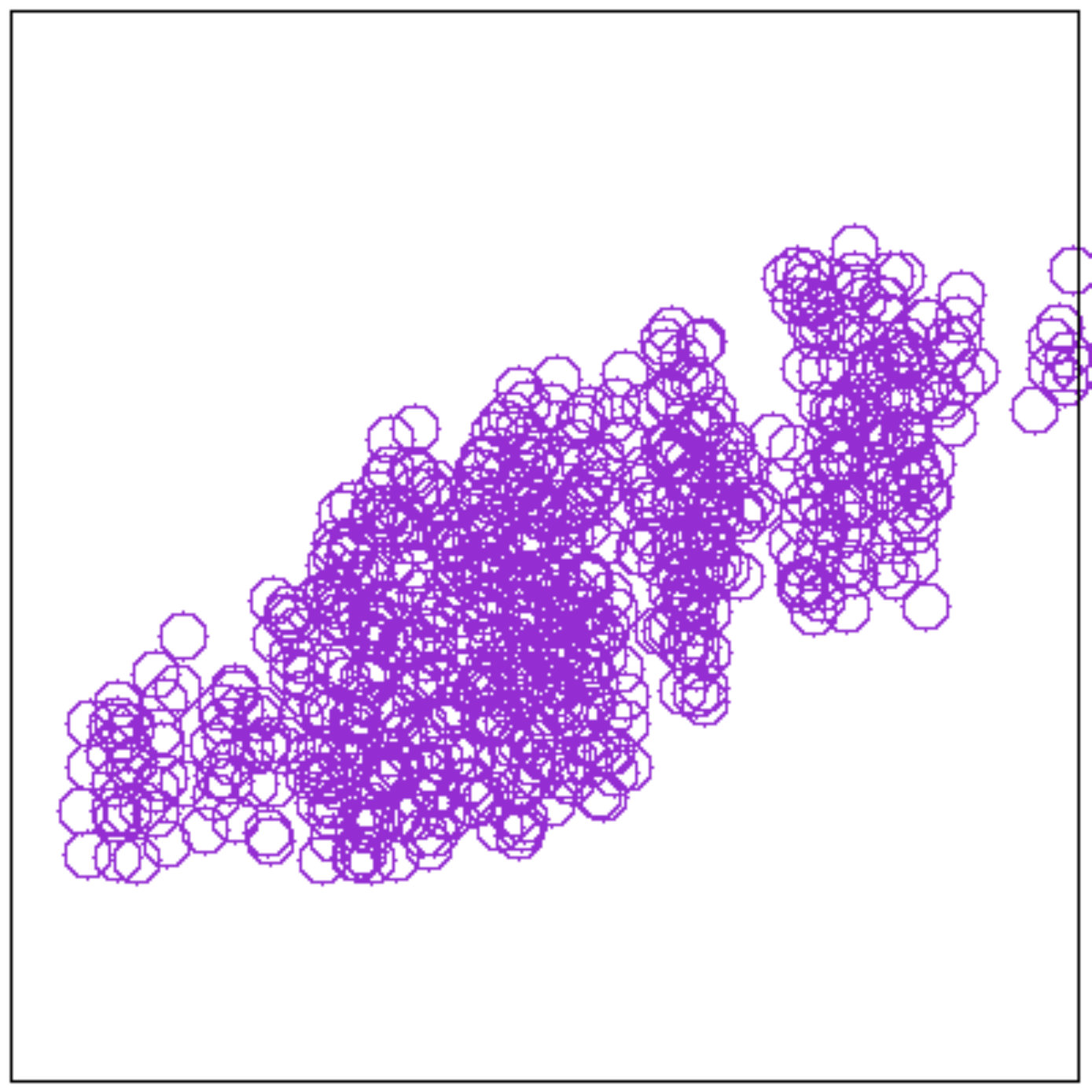}{0.3\textwidth}{$t = 17.7$ hours}
            \fig{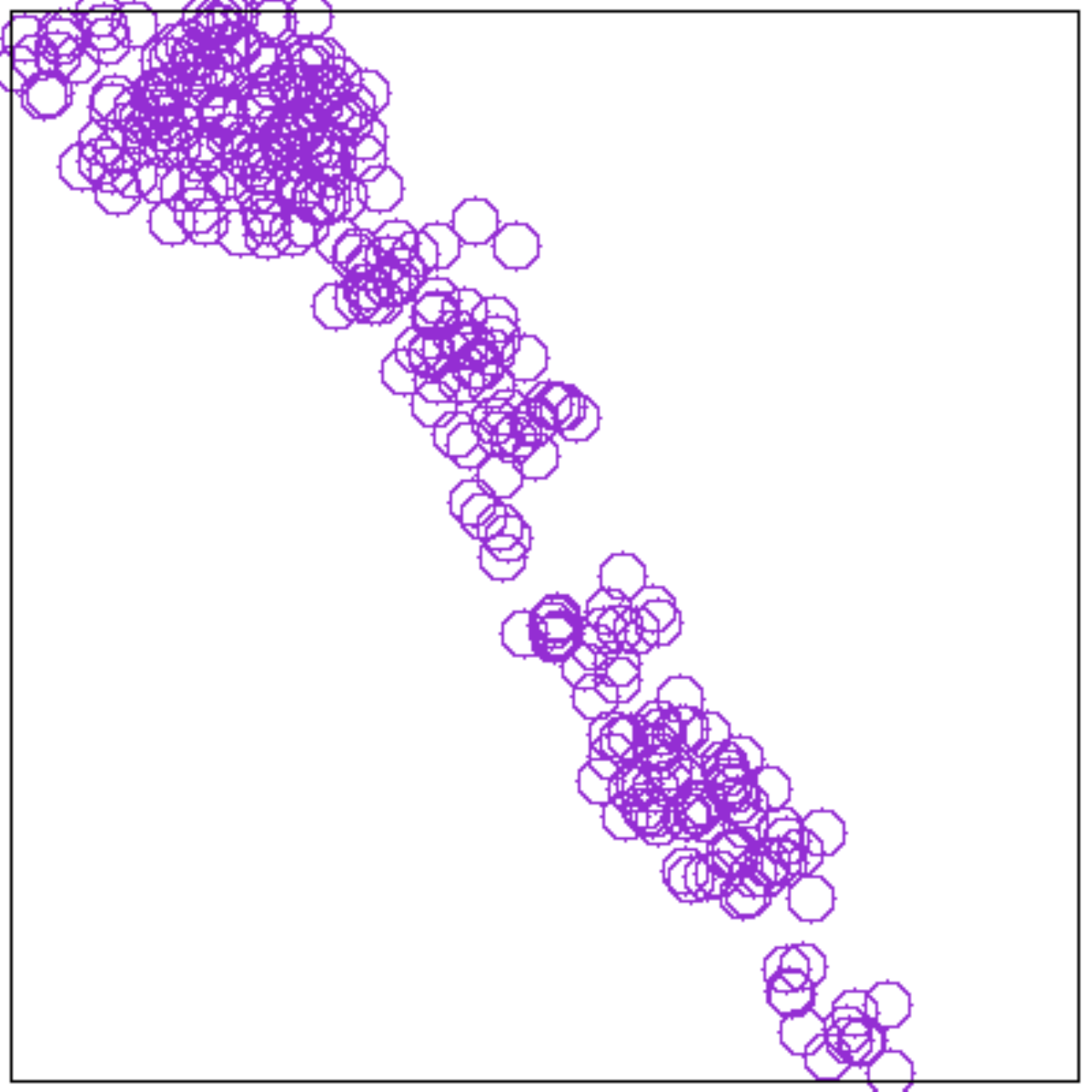}{0.3\textwidth}{$t = 22.2$ hours}
            }
  \caption{A magnified view of the rubble pile object destroyed by tidal force shown in Fig. \ref{fig:rochedisruptorbit}. The size of the circular marker corresponds to the size of each particle that composes the rubble pile object.
  \label{fig:rochedisruptrel}}
\end{figure*}

Figures \ref{fig:rochedisruptorbit} and \ref{fig:rochedisruptrel} show examples of the tidal destruction of a rubble pile object. In these examples, the rubble pile object deformed over time and resulted in accelerated tidal destruction. When a tidal break occurred, the dispersed particles did not reaccumulate. When the tidal destruction did not occur, the rubble pile object was observed to be tide-deformed and stretched in an ellipsoidal manner toward the central object.

\begin{singlespace}
\begin{table*}
  \caption{A list of the final states of the rubble pile object in each executed calculation. The parameters $\epsilon_n$ and $\epsilon_t$ are the coefficients of restitution and $a_{\mathrm{mean}}$ is the time average of the orbital major radius of the rubble pile object. “D” means that the rubble pile object has been destroyed (disrupted) and “ND” means that it has not been destroyed (non-disrupted).}
  \footnotesize
  \label{table:ifdisrupt}
  \begin{center}
	  \begin{tabular}{ccccc|ccccc|ccccc}
						\hline
					  Run
						&
					  $\epsilon_n$
					  &
					  $\epsilon_t$
					  &
					  \shortstack{$a_{\mathrm{mean}}$\\$[r_{\mathrm{ring}}]$}
					  &
					  D or ND
					  &
					  Run
						&
					  $\epsilon_n$
					  &
					  $\epsilon_t$
					  &
					  \shortstack{$a_{\mathrm{mean}}$\\$[r_{\mathrm{ring}}]$}
					  &
					  D or ND
					  &
					  Run
						&
					  $\epsilon_n$
					  &
					  $\epsilon_t$
					  &
					  \shortstack{$a_{\mathrm{mean}}$\\$[r_{\mathrm{ring}}]$}
					  &
					  D or ND \\
						\hline
						1 & 0.0 & 0.0 & 0.77 & D & 44 & 0.2 & 1.0 & 1.22 & ND & 87 & 0.8 & 0.0 & 0.97 & ND\\
						2 & 0.0 & 0.0 & 0.82 & D & 45 & 0.4 & 0.0 & 0.97 & D & 88 & 0.8 & 0.0 & 0.99 & ND\\
						3 & 0.0 & 0.0 & 0.83 & D & 46 & 0.4 & 0.0 & 0.99 & D & 89 & 0.8 & 0.2 & 0.94 & D\\
						4 & 0.0 & 0.0 & 0.84 & ND & 47 & 0.4 & 0.0 & 1.01 & ND & 90 & 0.8 & 0.2 & 0.97 & ND\\
						5 & 0.0 & 0.0 & 0.85 & ND & 48 & 0.4 & 0.0 & 1.04 & ND & 91 & 0.8 & 0.2 & 0.99 & ND\\
						6 & 0.0 & 0.2 & 1.04 & D & 49 & 0.4 & 0.2 & 0.95 & D & 92 & 0.8 & 0.4 & 1.02 & D\\
						7 & 0.0 & 0.2 & 1.08 & D & 50 & 0.4 & 0.2 & 0.97 & ND & 93 & 0.8 & 0.4 & 1.03 & D\\
						8 & 0.0 & 0.2 & 1.11 & ND & 51 & 0.4 & 0.2 & 1.00 & ND & 94 & 0.8 & 0.4 & 1.06 & ND\\
						9 & 0.0 & 0.2 & 1.14 & ND & 52 & 0.4 & 0.4 & 0.95 & D & 95 & 0.8 & 0.4 & 1.09 & ND\\
						10 & 0.0 & 0.4 & 0.99 & D & 53 & 0.4 & 0.4 & 0.98 & ND & 96 & 0.8 & 0.6 & 1.13 & D\\
						11 & 0.0 & 0.4 & 1.02 & D & 54 & 0.4 & 0.4 & 1.00 & ND & 97 & 0.8 & 0.6 & 1.15 & D\\
						12 & 0.0 & 0.4 & 1.05 & D & 55 & 0.4 & 0.6 & 0.95 & D & 98 & 0.8 & 0.6 & 1.17 & ND\\
						13 & 0.0 & 0.4 & 1.08 & ND & 56 & 0.4 & 0.6 & 0.97 & ND & 99 & 0.8 & 0.6 & 1.20 & ND\\
						14 & 0.0 & 0.6 & 0.85 & D & 57 & 0.4 & 0.6 & 1.00 & ND & 100 & 0.8 & 0.8 & 1.27 & D\\
						15 & 0.0 & 0.6 & 0.98 & D & 58 & 0.4 & 0.8 & 1.04 & D & 101 & 0.8 & 0.8 & 1.28 & D\\
						16 & 0.0 & 0.6 & 1.04 & ND & 59 & 0.4 & 0.8 & 1.07 & D & 102 & 0.8 & 0.8 & 1.31 & ND\\
						17 & 0.0 & 0.6 & 1.06 & ND & 60 & 0.4 & 0.8 & 1.08 & ND & 103 & 0.8 & 0.8 & 1.33 & ND\\
						18 & 0.0 & 0.8 & 0.85 & D & 61 & 0.4 & 0.8 & 1.10 & ND & 104 & 0.8 & 1.0 & 1.38 & D\\
						19 & 0.0 & 0.8 & 0.91 & D & 62 & 0.4 & 1.0 & 1.27 & D & 105 & 0.8 & 1.0 & 1.40 & D\\
						20 & 0.0 & 0.8 & 0.95 & ND & 63 & 0.4 & 1.0 & 1.29 & ND & 106 & 0.8 & 1.0 & 1.44 & ND\\
						21 & 0.0 & 0.8 & 0.97 & ND & 64 & 0.4 & 1.0 & 1.31 & ND & 107 & 0.8 & 1.0 & 1.49 & ND\\
						22 & 0.0 & 1.0 & 1.06 & D & 65 & 0.6 & 0.0 & 0.92 & D & 108 & 1.0 & 0.0 & 0.92 & D\\
						23 & 0.0 & 1.0 & 1.10 & D & 66 & 0.6 & 0.0 & 0.93 & ND & 109 & 1.0 & 0.0 & 0.97 & D\\
						24 & 0.0 & 1.0 & 1.12 & ND & 67 & 0.6 & 0.0 & 0.95 & ND & 110 & 1.0 & 0.0 & 0.99 & ND\\
						25 & 0.2 & 0.0 & 1.00 & D & 68 & 0.6 & 0.2 & 0.93 & D & 111 & 1.0 & 0.0 & 1.02 & ND\\
						26 & 0.2 & 0.0 & 1.02 & D & 69 & 0.6 & 0.2 & 0.95 & ND & 112 & 1.0 & 0.2 & 1.02 & D\\
						27 & 0.2 & 0.0 & 1.06 & ND & 70 & 0.6 & 0.2 & 0.97 & ND & 113 & 1.0 & 0.2 & 1.03 & D\\
						28 & 0.2 & 0.2 & 1.03 & D & 71 & 0.6 & 0.4 & 0.95 & D & 114 & 1.0 & 0.2 & 1.06 & ND\\
						29 & 0.2 & 0.2 & 1.05 & D & 72 & 0.6 & 0.4 & 0.97 & ND & 115 & 1.0 & 0.2 & 1.08 & ND\\
						30 & 0.2 & 0.2 & 1.07 & ND & 73 & 0.6 & 0.4 & 0.99 & ND & 116 & 1.0 & 0.4 & 1.10 & D\\
						31 & 0.2 & 0.2 & 1.09 & ND & 74 & 0.6 & 0.6 & 1.03 & D & 117 & 1.0 & 0.4 & 1.11 & D\\
						32 & 0.2 & 0.4 & 0.99 & D & 75 & 0.6 & 0.6 & 1.04 & D & 118 & 1.0 & 0.4 & 1.13 & ND\\
						33 & 0.2 & 0.4 & 1.01 & D & 76 & 0.6 & 0.6 & 1.05 & ND & 119 & 1.0 & 0.4 & 1.15 & ND\\
						34 & 0.2 & 0.4 & 1.04 & ND & 77 & 0.6 & 0.6 & 1.07 & ND & 120 & 1.0 & 0.6 & 1.25 & D\\
						35 & 0.2 & 0.6 & 0.95 & D & 78 & 0.6 & 0.8 & 1.19 & D & 121 & 1.0 & 0.6 & 1.27 & D\\
						36 & 0.2 & 0.6 & 0.96 & D & 79 & 0.6 & 0.8 & 1.21 & ND & 122 & 1.0 & 0.6 & 1.31 & ND\\
						37 & 0.2 & 0.6 & 0.99 & ND & 80 & 0.6 & 0.8 & 1.23 & ND & 123 & 1.0 & 0.8 & 1.33 & D\\
						38 & 0.2 & 0.8 & 0.95 & D & 81 & 0.6 & 1.0 & 1.32 & D & 124 & 1.0 & 0.8 & 1.38 & ND\\
						39 & 0.2 & 0.8 & 0.96 & D & 82 & 0.6 & 1.0 & 1.34 & D & 125 & 1.0 & 0.8 & 1.43 & ND\\
						40 & 0.2 & 0.8 & 0.97 & ND & 83 & 0.6 & 1.0 & 1.37 & ND & 126 & 1.0 & 1.0 & 1.60 & D\\
						41 & 0.2 & 0.8 & 0.99 & ND & 84 & 0.6 & 1.0 & 1.39 & ND & 127 & 1.0 & 1.0 & 2.05 & D\\
						42 & 0.2 & 1.0 & 1.19 & D & 85 & 0.8 & 0.0 & 0.93 & D & 128 & 1.0 & 1.0 & 2.38 & D\\
						43 & 0.2 & 1.0 & 1.20 & D & 86 & 0.8 & 0.0 & 0.94 & D & 129 & 1.0 & 1.0 & 2.82 & D\\
					  \hline
	  \end{tabular}
  \end{center}
  \normalsize
\end{table*}
\end{singlespace}

A list of the simulation results is shown in Table \ref{table:ifdisrupt}. For any coefficient of restitution parameters other than ($\epsilon_n$, $\epsilon_t$) = (1.0, 1.0), tidal destruction occurs (“D”) if the major mean orbital radius $a_{\mathrm{mean}}$ is smaller than a certain value, while tidal destruction does not occur (“ND”) if $a_{\mathrm{mean}}$ is larger than a certain value. For ($\epsilon_n$, $\epsilon_t$) = (1.0, 1.0), tidal disruption occurred in all simulations and a major orbital radius where the object was not tidal destroyed could not be found.

\begin{table*}
  \caption{The position of the Roche radius for each set of restitution coefficient parameters ($\epsilon_n$, $\epsilon_t$). Values are normalized by $r_{\mathrm{ring}}$.}
  \label{table:Rochelimit}
  \begin{center}
	  \begin{tabular}{r|cccccc}
		  \hline
		  \ & $\epsilon_t$ = 0.0 & 0.2 & 0.4 & 0.6 & 0.8 & 1.0 \\
		  \hline \hline
		  $\epsilon_n$ = 0.0 & 0.83 & 1.10 & 1.06 & 1.01 & 0.93 & 1.11 \\
		  0.2 & 1.04 & 1.06 & 1.02 & 0.98 & 0.96 & 1.21 \\
		  0.4 & 1.00 & 0.96 & 0.97 & 0.96 & 1.07 & 1.28 \\
		  0.6 & 0.92 & 0.94 & 0.96 & 1.04 & 1.20 & 1.36 \\
		  0.8 & 0.96 & 0.96 & 1.05 & 1.16 & 1.30 & 1.42 \\
		  1.0 & 0.98 & 1.04 & 1.12 & 1.29 & 1.35 & $>$2.82 \\

					  \hline
	  \end{tabular}
  \end{center}
\end{table*}

For each restitution coefficient ($\epsilon_n$, $\epsilon_t$), the maximum orbital major radius that causes tidal destruction $a_{D,\mathrm{max}}$, and the minimum orbital major radius that does not cause tidal destruction $a_{ND,\mathrm{min}}$ are calculated. In the case of ($\epsilon_n$, $\epsilon_t$) = (1.0, 1.0), the orbital major radius that would not cause tidal destruction was not found; therefore, the Roche radius was assumed to be more than $a_{D,\mathrm{max}} = 2.82$ $r_{\mathrm{ring}}$. A list of the Roche radii obtained in this way are shown in Table \ref{table:Rochelimit}. The roche radius of any restitution coefficient parameter was greater than the rigid-body roche radius, $r_{\mathrm{roche,rigid}} = 0.71$ $r_{\mathrm{ring}}$. Furthermore, for many restitution coefficient parameters, the roche radius was smaller than the perfect fluid roche radius, $r_{\mathrm{roche,fluid}} = 1.09$ $r_{\mathrm{ring}}$. Therefore, for most coefficients of restitution parameters, the Roche radius was near the ring radius. This is consistent with our scenario in which the formation of Haumea's ring was formed just inside the Roche radius.

However, for some parameters with large $\epsilon_n$ and $\epsilon_t$, the roche radius was larger than $r_{\mathrm{roche,fluid}}$; this parameter space is considered outside the scope of our numerical calculations. This may be due to the numerical calculation model for the rubble pile object. The model of the rubble pile object in this study ignores adhesive force and thus may be more easily destroyed than the actual object.

\subsection{Discussions}

In this calculation, the density of the rubble pile is assumed to be that of ice; however, the actual density of the particles that make up the ring has not been determined. In general, the Roche radius is proportional to 1/3 of the power of the density of an object revolving around a central celestial body. Therefore, if the density of the ring particles is different, the position of the corresponding Roche radius will also be slightly different. However, the dependence on the 1/3 of power is not very high, so it is unlikely that the difference in density will have a significant effect on the ring formation scenario in this study.

A more detailed model needs to be applied to accurately investigate the tidal destruction process of the rubble pile. The tidal disruption distance not only depends on density, material strength and other material properties, but also shape and rotation period of the companion, and it even depends on its internal structure (whether it is differentiated or not). We should consider the distance from the primary at which a disk of particles would not be able to accrete to form a single body. We focused on only the disruption of rubble pile objects in our study and did not follow a long-term evolution of the disrupted particles. An important future work is determining the distance at which the tidal force would not prevent a satellite from forming by accretion of the disk material \citep{2012Sci...338.1196C}.

\section{Conclusion}

In this study, we present a ring formation scenario for the dwarf plant Haumea modelled after \citet{2012MNRAS.419.2315O}. To verify this scenario, the position of the orbital unstable region boundary caused by Haumea's non-axisymmetric gravitational field and the position of the Roche radius were obtained using $N$-body calculations.

Calculation of the orbit of a particle revolving near Haumea revealed that the particle was removed from the region inside the 2:1 orbital resonance on a time scale of less than 10 days. This is consistent with the result of \citet{2019NatAs...3..146S}, in which the orbit of a particle becomes unstable because of strong torque due to the 2:1 orbital resonance and the Lindblad resonance inside it. The boundary of the final orbital instability region was 0.905 $r_{\mathrm{ring}}$, just inside the current ring of Haumea.

Analytical calculation of the position of the Roche radius for a rigid body revolving around Haumea yielded a position of 0.71 $r_{\mathrm{ring}}$. When Haumea was regarded as a sphere of the same volume, the position of the perfect fluid roche radius was 1.09 $r_{\mathrm{ring}}$. This result is consistent with the model used in this study in which a ring had formed just inside the roche radius, considering that the roche radius for a real object is somewhere between the rigid body and the fluid roche radii.

As a result of the $N$-body calculation of the position of the Roche radius for the object revolving around Haumea, the roche radius became close to Haumea's current ring with many parameters of the restitution coefficient.  Furthermore, the numerically determined values of the Roche radius are consistent with the analytically determined values of the Roche radius for many parameters.

In summary, it can be concluded that the position of the boundary of an unstable orbit region is likely just inside Haumea's current ring position; the position of the Roche radius is likely to be just outside it. In other words, it is inevitable that Haumea's ring will be located near the 3:1 orbital resonance, among which the ring was actually found.

Haumea's ring formation scenario in this study can provide suggestions for the formation of other celestial ring systems. In this calculation of the boundary of orbital unstable regions, it is considered that the position is determined by 2:1 orbital resonance. This resonance moves inward if the central celestial body rotates fast and outward if it rotates slow. However, the position of the Roche radius does not depend greatly on the rotation period and shape of the central celestial body. Therefore, expected characteristics of ring systems surrounding celestial bodies exhibiting a non-axisymmetric shape can be classified as follows. (1) For a celestial body whose rotation period is shorter than that of Haumea, the ring can exist in the wide area inside the Roche radius because the boundary of the unstable region is closer to the central body than that of Haumea. (2) If a celestial body had the same rotation period as Haumea, the ring could exist only in the narrow area because the boundary of the unstable area and the position of the Roche radius are close. (3) An object with a rotation period longer than that of Haumea is unlikely to have a ring because the boundary of the unstable region is outside the Roche radius. If a ring is found in a non-axisymmetric object other than Haumea via future occultation observations of TNO, it may be possible to evaluate the universality of the ring formation scenario presented in this study.

\acknowledgments

We gratefully acknowledge the reviewer's helpful comments. This work was supported by JSPS KAKENHI grant number 19K03950.

\appendix

\section{Gravitational field potential around a triaxial ellipsoid}\label{sec:appendix_a}

An equation representing the shape of a general three-axis unequal ellipsoid is given here:
\begin{equation}
\frac{x^2}{a^2}+\frac{y^2}{b^2}+\frac{z^2}{c^2}=1 \ \ \ (a>b>c).
\end{equation}
According to \citet{kellogg_1955}, the gravitational potential $U (x, y, z)$ at a point $(x, y, z)$ outside the triaxial spheroid is given by:
\begin{equation}
	U(x,y,z)=2\pi G abc\rho_M \left[ D(\lambda) - A(\lambda)x^2 - B(\lambda)y^2 - C(\lambda)z^2 \right],
	\label{eq:U}
\end{equation}
where
\begin{equation}
	A(\lambda)=\frac{2}{(a^2 - c^2)^{3/2} k^2}\left[ F(k,\phi ) - E(k,\phi ) \right],
\end{equation}
\begin{equation}
	B(\lambda)=\frac{2}{(a^2 - c^2)^{3/2} k^2 k'^2}\left[ E(k,\phi ) - k'^2 F(k,\phi ) k^2 \frac{\sin \phi \cos \phi}{\sqrt{1-k^2 \sin^2 \phi}} \right],
\end{equation}
\begin{equation}
	C(\lambda)=\frac{2}{(a^2 - c^2)^{3/2} k'^2}\left[ \frac{\sin \phi \sqrt{1-k^2 \sin^2 \phi}}{\cos \phi} - E(k,\phi ) \right],
\end{equation}
\begin{equation}
	D(\lambda)=\frac{2}{(a^2 - c^2)^{1/2}}F(k,\phi ),
\end{equation}
and
\begin{equation}
	\sin \phi = \sqrt{\frac{a^2 -c^2}{a^2 + \lambda}} \ \ \ (0<\phi < \pi /2),
	\label{eq:phi}
\end{equation}
\begin{equation}
	k=\sqrt{\frac{a^2 - b^2}{a^2 - c^2}},
\end{equation}
\begin{equation}
	k'=\sqrt{\frac{b^2 - c^2}{a^2 - c^2}}.
\end{equation}
$\lambda$ is the largest real root of the following cubic equation,
\begin{equation}
	f(s)=\frac{x^2}{a^2 + s}+\frac{y^2}{b^2 + s}+\frac{z^2}{c^2 + s}-1=0.
	\label{eq:fs}
\end{equation}
$F$ and $E$ are incomplete elliptic integrals of the first and second kind,
\begin{equation}
	F(k,\phi)=\int_0^{\phi} \frac{\mathrm{d}\phi}{\sqrt{1-k^2 \sin^2 \phi}}
	\label{eq:F}
\end{equation}
\begin{equation}
	E(k,\phi)=\int_0^{\phi} \sqrt{1-k^2 \sin^2 \phi} \ \mathrm{d}\phi.
	\label{eq:E}
\end{equation}

\section{Tidal force from a homogeneous ellipsoid} \label{sec:appendix_b}

According to Equation (\ref{eq:U}), the gravitational potential of the central object at the position of the companion, $(x, 0, 0)$, is
\begin{equation}
	U(x,0,0)=2\pi G abc\rho_M \left[ D(\lambda) - A(\lambda)x^2 \right],
\end{equation}
where $\lambda = x^2 - a^2$ with $y = z = 0$ at Equation (\ref{eq:fs}). Thus, from Equation (\ref{eq:phi}),
\begin{equation}
	\phi = \arcsin \sqrt{\frac{a^2 - c^2}{x^2}}.
\end{equation}
Therefore, Equations (\ref{eq:F}) and (\ref{eq:E}) can be written as a function of $x$ in:
\begin{equation}
	F(x)=\int_0^{\arcsin \sqrt{\frac{a^2 - c^2}{x^2}}} \frac{\mathrm{d}\phi}{\sqrt{1-k^2 \sin^2 \phi}},
\end{equation}
\begin{equation}
	E(x)=\int_0^{\arcsin \sqrt{\frac{a^2 - c^2}{x^2}}} \sqrt{1-k^2 \sin^2 \phi} \ \mathrm{d}\phi.
\end{equation}
Thus,
\begin{equation}
	\frac{dF(x)}{dx}=-\sqrt{\frac{(x^2 - k^2 (a^2 - c^2))(a^2 - c^2)}{x^2 - (a^2 - c^2)}},
\end{equation}
\begin{equation}
	\frac{dE(x)}{dx}=-\frac{1}{x^2}\sqrt{\frac{(x^2 - k^2 (a^2 - c^2))(a^2 - c^2)}{x^2 - (a^2 - c^2)}}.
\end{equation}
The $x$ component of gravity from the central celestial body acting on the companion at position $(x, 0, 0)$ is:
\begin{equation}
	\mathcal{F}_x (x,0,0)=-\frac{d}{dx}U(x,0,0) = -4 \pi G abc\rho_M \frac{F(x)-E(x)}{(a^2 - c^2)^{3/2} k^2}.
\end{equation}
Therefore,
\begin{equation}
	\mathcal{F}_T = r\frac{d}{dx}\mathcal{F}_x (x,0,0) = -\frac{4 \pi G abc\rho_M r}{(a^2 - c^2)^{3/2} k^2} \left( F(x)-E(x) + x \frac{d}{dx}(F(x)-E(x))\right).
\end{equation}

\bibliographystyle{aasjournal}

\end{document}